\documentclass[10pt, preprint]{emulateapj}

\usepackage{color}
\usepackage{amsmath}	

\usepackage{style_wei_symbols}
\usepackage{style_wei}

\usepackage{natbib}
\begin{document}

\title{Combined Modeling of Acceleration, Transport, and Hydrodynamic Response in Solar Flares: I.~The Numerical Model}


\author{Wei Liu\altaffilmark{1}\altaffilmark{2}, Vah\'{e} Petrosian\altaffilmark{2}, 
and John T. Mariska\altaffilmark{3}}


\altaffiltext{1}{Stanford-Lockheed Institute for Space Research, 466 Via Ortega, Cypress Hall, 
 Stanford, CA 94305-4085}
\altaffiltext{2}{Department of Physics, Stanford University,	
 Stanford, CA 94305-4060}
\altaffiltext{3}{Naval Research Laboratory, Code 7673, Washington, DC 20375-5000}


\shorttitle{Combined Modeling of Acceleration, Transport, and Hydrodynamics in Solar Flares} 
\shortauthors{Liu et al.}
\slugcomment{{\it ApJ 2009, in Press; Received 2009 April 9; Accepted 2009 June 12}}

\begin{abstract}	

Acceleration and transport of high-energy particles and fluid dynamics of atmospheric plasma
are interrelated aspects of solar flares, but for convenience and simplicity they
were artificially separated in the past. We present here
self-consistently combined Fokker-Planck modeling of particles and 
hydrodynamic simulation of flare plasma.	
Energetic electrons are modeled with the Stanford unified code of acceleration, transport, 
and radiation, while plasma is modeled with the Naval Research Laboratory flux tube code. 
We calculated the collisional heating rate directly from the particle transport code, 
which is more accurate than those in previous studies based on approximate analytical solutions.
  We repeated the simulation of \citet{MariskaJ1989ApJ...341.1067M} with an injection
of power-law, downward beamed electrons using the new heating rate.
For this case, a $\sim$10\% difference was found from their old result.
  We also used a realistic spectrum of injected electrons provided by the stochastic acceleration
model, which has a smooth transition from a quasi-thermal background at low energies to a nonthermal tail
at high energies.
  The inclusion of low-energy electrons results in relatively more heating in the
corona (vs.~chromosphere) and thus a larger downward heat conduction flux. The interplay 
of electron heating, conduction, and radiative loss leads to stronger chromospheric evaporation
than obtained in previous studies, which had a deficit in low-energy electrons due to
an arbitrarily assumed low-energy cutoff.
The energy and spatial distributions of energetic electrons and bremsstrahlung photons bear signatures
of the changing density distribution caused by chromospheric evaporation.
In particular, the density jump at the evaporation front gives rise to enhanced emission,
which, in principle, can be imaged by X-ray telescopes.
 This model can be applied to investigate a variety of high-energy processes 
in solar, space, and astrophysical plasmas.
\end{abstract}

\keywords{acceleration of particles---hydrodynamics---methods: numerical---
Sun: chromosphere---Sun: flares---Sun: X-rays, gamma rays}	

\section{Introduction}
\label{sect_intro}

A solar flare, as one of the most prominent manifestations of solar activity,
has many faces among which are acceleration and transport of high-energy particles and the 
dynamical response of atmospheric plasma. 
It is generally believed that magnetic reconnection in the corona is the primary energy release
mechanism that leads to plasma heating and particle acceleration. The heated plasma and accelerated 
particles (primarily electrons) produce bremsstrahlung X-rays at the apex of the flare loop observed 
as a loop-top (LT) source \citep[e.g.,][]{Masuda1994Nature, 
PetrosianV2002ApJ...569..459P, KruckerS.occult-stat2008ApJ...673.1181K}.
Some of the released energy is transported down the		
{\it closed} magnetic loop by nonthermal particles (electrons and ions) and thermal conduction,
which contribute to energy gain in various layers of the atmosphere.
Electrons give up most of their energy to ambient particles via Coulomb collisions,
and produce hard X-rays (HXRs) primarily at the footpoints (FPs) of the loop in the dense transition region (TR) 
and chromosphere \citep[see][]{HoyngP1981ApJ246L.155.HXRfp, SakaoT1994PhDT.......335S, Saint-Hilaire2008SolPh.FPasym}.
Accelerated protons and heavier ions, on the other hand, cause nuclear reactions while colliding
with background particles and produce $\gamma$-rays.
Some accelerated electrons and ions escape along {\it open} magnetic field lines into
the interplanetary space and are observed as solar energetic particles (SEPs) 
by {\it in situ} instruments \citep[e.g.,][]{LinRP.SEP-e.1985SoPh..100..537L, 
LiuS2004ApJ...613L..81L, KruckerS_e-spec-WIND_2007ApJ...663L.109K}.
The rate of energy gain in the chromosphere, if exceeding the combined local radiative and conductive cooling rate, 
can rapidly heat the plasma up to a temperature of $\sim$$10^7$ K.  The resulting overpressure drives an
upward mass flow at a speed up to hundreds of km~s$^{-1}$, which fills
the flare loop with a hot plasma, giving rise to the gradual increase of soft
X-ray (SXR) emission. This process, termed chromospheric evaporation by
\citet{NeupertW1968ApJ...153L..59N}, can influence particle transport by changing
the ambient density in the loop on timescales of tens of seconds, and 
affect heat conduction by changing the loop temperature distribution at the same time.
Collisional and conductive heating will be consequently modified, 
and in turn, so will the dynamic atmospheric response. 
On longer timescales of minutes, as magnetic reconnection proceeds and new loops
are formed or excited, the above processes repeat sequentially in newer loops.

\subsection{Motivation for This Study}	
\label{subsect_intro1}

The aforementioned processes are coupled in a circular chain, but due to great
complexity of the subject, previous researchers tended to 	
focus on one process at a time 
while assuming some simple forms for others. Past investigations	
fall into two general categories: (1) acceleration and transport of particles
and (2) fluid dynamics of atmospheric plasma. 

 (1) 
Various mechanisms have been proposed for particle acceleration. Among the agents of
acceleration are DC electric fields \citep{HolmanG1985ApJ...293..584H, LitvinenkoY.acc-in-CS.1996ApJ...462..997L, 
ZharkovaV.asym-acc.2004ApJ...604..884Z}, shocks \citep{TsunetaS1998ApJ...495L..67T},
and turbulence or plasma waves \citep{RamatyR1979AIPC...56..135R, HamiltonR1992ApJ...398..350H, 
MillerJ1996ApJ...461..445M, PetrosianV2004ApJ...610..550P}.
Particle transport is comparatively better understood and previous studies usually assumed 
a hydrostatic atmosphere. Early analytical studies 
\citep{BrownJ1973SoPh...31..143B, PetrosianV1973ApJ...186..291P, 
LinR.HudsonH.e-1976SoPh...50..153L, Emslie1978ApJ...224..241E,
Chandrashekar.Emslie.heating1987SoPh..107...83C} 
took various approximations (e.g., neglecting pitch-angle diffusion)
to allow the problem to be tractable. The numerical study of \citet{LeachJ1981ApJ...251..781L} improved on this by
solving the Fokker-Planck transport equation with inclusion of pitch-angle changes due to
Coulomb collision and magnetic mirroring. 
This was later extended to the relativistic regime by including energy losses and pitch-angle changes	
due to synchrotron radiation \citep{McTiernanJ1990ApJ...359..524M}. 
Similar Fokker-Planck studies of particle transport were performed by \citet{MacKinnonA1991A&A...251..693M},
\citet{McClementsK1992A&A...258..542M}, and \citet{SyniavskiiD1994ApJS...90..729S}. 

(2) Fluid dynamics of the magnetized atmosphere in response to flare heating can be
best studied with a multi-dimensional magnetohydrodynamic (MHD) model, 
but for simplicity most efforts were invested in one-dimensional hydrodynamic (HD) simulations.
This is justified for the solar corona where the magnetic pressure dominates the gas pressure (low $\beta$ plasma) 
and resistivity is low. As a result, plasma is only allowed to flow along the magnetic field lines
due to the line-tying condition. Previous HD models 
\citep{MacNeiceP.HD1984SoPh...90..357M, NagaiF1984ApJ...279..896N, EmslieNagai1985ApJ...288..779E,
FisherG1985ApJ...289..434F, FisherG1985ApJ...289..425F, FisherG1985ApJ...289..414F, 
MariskaJ1989ApJ...341.1067M, GanW1990ApJ...358..328G, Emslie.Li.Mariska1992ApJ...399..714E}
usually assumed a power-law spectrum of accelerated electrons injected at the apex of the loop, 
and calculated collisional heating along the loop by these electrons	
from approximate analytical solutions of particle transport mentioned above.	
\citet{AbbettW1999ApJ...521..906A} and \citet{AllredJ2005ApJ...630..573A} improved on previous
studies by including detailed calculation of radiative transfer in the atmosphere.
%

There are theoretical and observational motivations to investigate the particle and fluid aspects of a solar
flare together in a self-consistent manner.	
From a theoretical point of view, such an investigation is demanded
in order to retrieve missing physics when the two aspects were studied separately.
It has also become technically more feasible,	
thanks to advances in both aspects over the last three decades and particularly in recent years.
Several studies \citep{MillerJ2005AGUSMSP41C..02M, WinterH2007ASPC..369..501W} along this 
direction are already under way, but none of them has been completed.
%
From an observational point of view, new observations, particularly X-ray images
and spectra obtained by the current {\it Ramaty High Energy Solar Spectroscopic Imager} ({\it RHESSI}) and previous 
{\it Yohkoh} missions, have posed new challenges to the existing theories.
For example, in recent studies of the \citet{NeupertW1968ApJ...153L..59N} effect,
\citet{VeronigA2005ApJ...621..482V} and \citet{LiuW2006ApJ...649.1124L} found that, 
unexpectedly, the electron energy deposition power, which is more physically related
to the thermal energy change rate, did not yield a better correlation with the time derivative
of the SXR flux than the conventional HXR flux.
In an event of chromospheric evaporation imaged by {\it RHESSI} for the first time,
\citet{LiuW2006ApJ...649.1124L} found X-ray sources
moving from the FPs to the LT at very high speeds ($\sim$10$^3$~km~s$^{-1}$). 
More interestingly, \citet{SuiL2006ApJ...645L.157S} found double nonthermal sources moving first downward from the LT
toward the FPs and then upward along the loop.  To fully understand these observations requires
a joint study of acceleration and transport of particles and fluid dynamics of
the atmospheric response.

\subsection{Approach of This Study}
\label{subsect_intro2}

With the goal to investigate the coupled processes of acceleration, transport, and hydrodynamics 
in solar flares, we present here combined Fokker-Planck modeling of particles and HD simulation of plasma.
 (1) The Fokker-Planck model utilizes the Stanford unified code of 
particle acceleration, transport, and bremsstrahlung radiation \citep{PetrosianV2001flareCode}. 
The transport and radiation calculation is based on the work of
\citet{LeachJ1981ApJ...251..781L, LeachJ1983ApJ...269..715L} and \citet{McTiernanJ1990ApJ...359..524M}.
The acceleration module of the code adopts the stochastic acceleration
model of \citet[][hereafter {\bf PL04}]{PetrosianV2004ApJ...610..550P}, which has inherited knowledge 
accumulated over a decade \citep{HamiltonR1992ApJ...398..350H, DungPetrosian1994ApJ,
ParkB1995ApJ...446..699P, ParkB1996ApJS..103..255P, ParkB1997ApJ...489..358P}.
When compared with observations, this model has many attractive features and 
advantages \citep{LiuS2004ApJ...613L..81L, LiuS2006ApJ...636..462L, LiuW_2LT.2008ApJ...676..704L} 
over other mechanisms. 
 (2) The HD simulation uses the Naval Research Laboratory (NRL) Solar Flux Tube Model  
\citep*[][hereafter {\bf MEL89}]{MariskaJ1989ApJ...341.1067M}, which, 
as a modified version of the \citet{MariskaJ1982ApJ...255..783M} model,
provides excellent treatment of fluid dynamics and has been widely used in studying 
atmospheric response to flare heating \citep[e.g.,][]{WarrenHP.multi.2005ApJ...618L.157W}.

One of the major advances marked by this study is the more accurate and self-consistent evaluation
of the collisional heating rate by nonthermal electrons. This heating rate is critical to
HD simulation of flares, but was not properly calculated previously in two major aspects:
 (1) The calculation of energy loss of energetic electrons and thus the heating rate was based on 
approximate analytical solutions \citep[e.g.,][]{BrownJ1973SoPh...31..143B, Emslie1978ApJ...224..241E},
which incorporated only pitch angle {\it growth} due to Coulomb collisions,
but in reality the pitch angle change is a {\it diffusion} process.	
This will be remedied in this study with the inclusion of a full Fokker-Planck treatment of electron transport.
 (2) Another previous drawback was the use of an unrealistic spectrum of injected electrons, 
which usually was a power-law with a low-energy cutoff. 
\citet{FisherG1985ApJ...289..414F}, for example, assumed	
a sharp low-energy cutoff at $E_{\rm c} =20 \keV$ (i.e., no electrons below $E_{\rm c}$), while 
MEL89 introduced a ``soft" cutoff below which the spectrum is still a power-law with a positive slope.	
It should be noted that
rather than an intrinsic property of the primary accelerated electron population,
a low-energy cutoff or turnover	
in the electron spectrum inferred from X-ray observations can result from secondary effects such as return currents
\citep{ZharkovaV.return-current.2006ApJ...651..553Z} and photospheric albedo
\citep{Langer.Petrosian1977ApJ...215..666L, BaiT.albedo1978ApJ...219..705B, SuiL.flattening2007ApJ...670..862S}.
  The collisional heating rate is sensitive to the injected electron spectrum
and thus the use of an incorrect spectrum would make the HD simulation deviate from reality significantly.
PL04 has provided a more realistic electron spectrum that has a continuous span from a  
quasi-thermal distribution at low energies to a nonthermal tail at high energies, avoiding
unnecessary low-energy cutoff.  It also gives good fits to both
LT and FP X-ray spectra obtained by \hsiA. Such an electron spectrum is used in this work. As we will see later,
the low-energy electrons, which otherwise would have been missing if a cutoff were to be present,
play an important role in heating and in influencing the subsequent HD evolution.

We present the numerical model in \S~\ref{sect_model}, techniques to combine
different modules of the model in \S~\ref{sect_comb-codes}, and simulation runs in
\S~\ref{sect_result}. We compare the HD characteristics of different simulations in \S~\ref{sect_cmpr}
and examine the HD effects on particle transport and X-ray emission in \S~\ref{sect_el-ph}.
Conclusions and discussions are given in \S~\ref{sect_conclude}.
%
This model is based on the PhD thesis of \citet{LiuW2006PhDT........35L}, which has also appeared 
as a book \citep{LiuW.2008sfpa.book.....L}. The model has been refined ever since
and the numerical results presented here are new. Our new calculations and analyses include:
(1) recomputing the MEL89 simulation using our transport code (see \S\ref{subsect_RunH}),
(2) detailed analysis of the interplay of heating and cooling (\S\ref{subsect_heating}),
and (3) examining the temperature distribution of plasma velocity which can be directly
compared with Doppler observations (\S\ref{subsect_vel-distr}).
In order to achieve more accuracy, we have extended the number of pitch angle bins from 24 to 100.
 

\section{Simulation Model}
\label{sect_model}

\begin{figure}[thb]        
 \epsscale{1} 
 \plotone{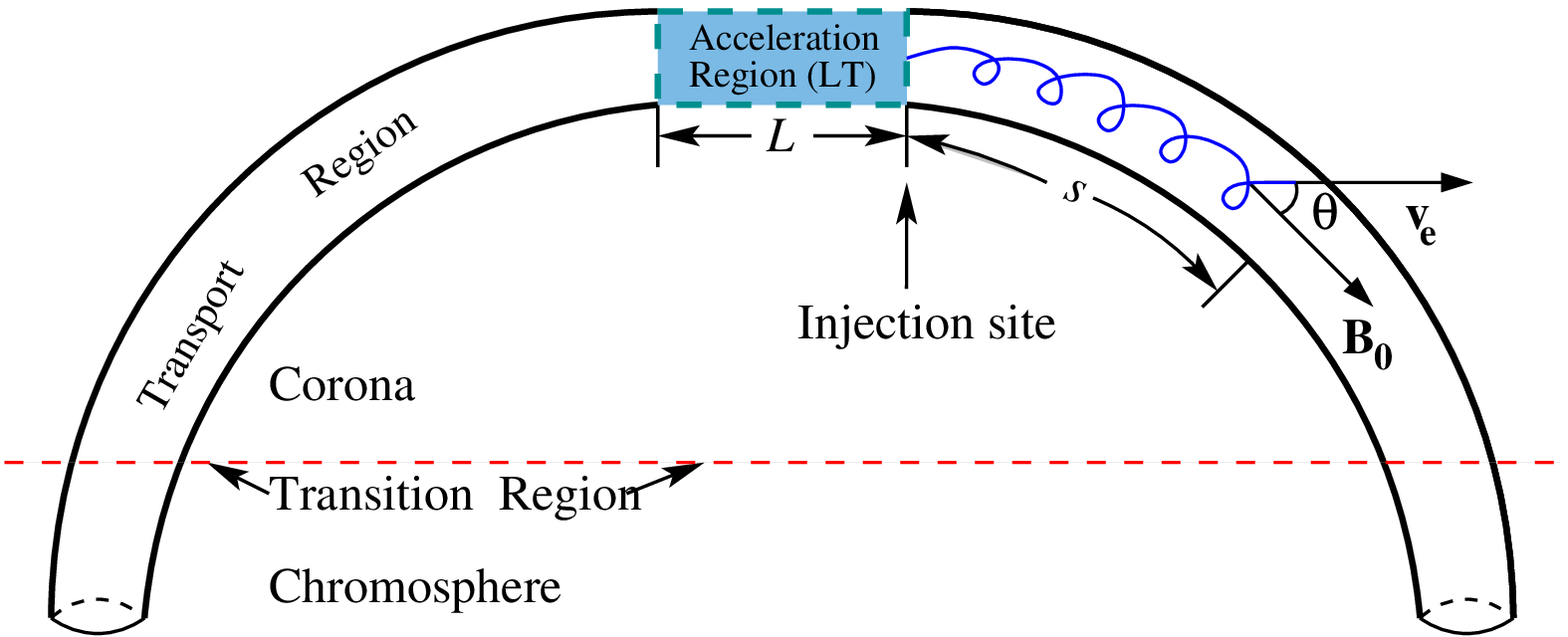} 
 \vskip 0.2in
 \plotone{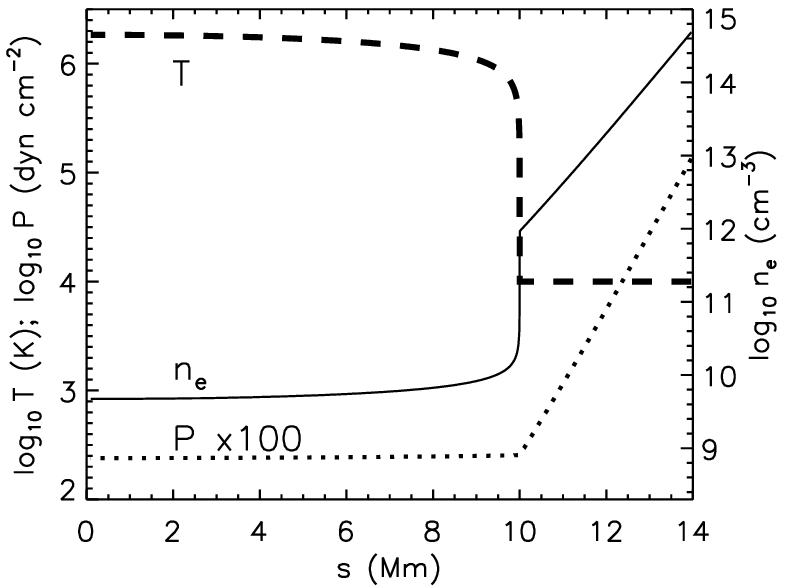} 
 \caption[Schematic of the model geometry]
 {  {\it Top:} Geometry of the model flare loop. $\theta$ is the pitch angle
 of the electron with a velocity ${\bf v_e}$ in the guiding magnetic field ${\bf B_0}$.
   {\it Bottom:} Initial distribution of logarithmic temperature $T$, gas pressure $P$ ({\it left scale}),
 and electron number density $n_e$ ({\it right scale}) vs.~distance in the
 transport region of the loop. Pressure is scaled upward by a factor of 100. 
 } \label{model-geometry.eps}
 \end{figure}

Here we consider the dynamical evolution of a single flare loop perpendicular to the solar surface. 
As shown in Figure~\ref{model-geometry.eps} ({\it top}), an {\it acceleration region} of 
length $L=5 \Mm$ is located at the top of the model loop, 
sandwiched between		
two symmetric quarter-circles called the {\it transport region} of length $s_{\rm max}=14 \Mm$.
The loop has a uniform circular cross-section $a(s)=a_0=\pi r^2$ with a constant radius of $r=0.3 \Mm$
at any distance $s$ from the edge of the acceleration region. 
In its initial state (Fig.~\ref{model-geometry.eps}, {\it bottom})
the loop spans from the hot ($T\gtrsim 10^6 \K$), tenuous corona to the cold ($T=10^4 \K$), 
dense chromosphere, with the TR (defined here as the lowest point where $T \geq 10^5 \K$) 
located at $s_{\rm tr} \simeq 10 \Mm$.
%

The simulation model includes four parts: (1) The {\it stochastic acceleration code} generates
a spatially averaged distribution (in energy) of high-energy electrons in the acceleration region.
This distribution is fed to the transport region where (2) the {\it transport code}
computes the electron distribution (in energy and pitch angle) and collisional heating rate as a function of distance
(depth), and (3) the {\it hydrodynamic code} simulates the atmospheric response to this heating.
Finally, (4) the {\it radiation code} calculates the corresponding bremsstrahlung emission
in the loop. Parts~(1), (2), and (4) are inherited from the Stanford unified particle code
\citep{PetrosianV2001flareCode} in which the acceleration module has been revised according to
PL04, while part (3) is adopted from the NRL flux tube code (MEL89). 
Details of the four parts are described in the following subsections.

Note that sequential energizing (or excitation) propagating from one flare loop 
to another has been observed as the apparent motions of HXR LT 	
\citep{GallagherP2002SoPh..210..341G, SuiL2003ApJ...596L.251S}	
and FP \citep{GrigisBenz2005ApJ, YangYH.FPstat.2009ApJ...693..132Y, LiuW_FPAsym_2009ApJ...693..847L} 
sources. 	
In our simulations, we set the duration of the impulsive phase to be 60~s.
This time, according to \citet{SchrijverC.HudsonH.Oct28.2006ApJ...650.1184S},
translates into an apparent FP speed of ${2 r \over 60 \s}= 10 \km \ps$ (where $r=0.3 \Mm$),
which is comparable to the observed values \citep{LiuW2004ApJ...611L..53L}.
Our single loop scenario is thus legitimate within the relevant timescale and can be viewed as 
an elementary process of sequential excitation of multiple loops. 
Evolution on longer timescales (say, $\gtrsim$100~s) involves multiple loops 
and can be studied by superimposing sequential single-loop simulations
\citep[e.g.,][]{WarrenHP.multiloop.2006ApJ...637..522W}.




\subsection{Stochastic Acceleration}	
\label{subsect_SAmodel}


The stochastic acceleration model of PL04 addresses electron and proton acceleration 
by plasma waves propagating parallel to the background magnetic field ${\bf B}_0$.
According to this model, large-scale turbulence or long-wavelength plasma waves are generated in 
the corona as a result of magnetic reconnection. The turbulence, cascading to smaller scales, heats plasma 
and accelerates particles in a region near the top of the flaring loop.
The heated plasma and accelerated electrons produce the observed thermal and nonthermal X-rays,
respectively, in the acceleration region or the LT source 
\citep{XuY.fwdfit2008ApJ...673..576X, LiuW_2LT.2008ApJ...676..704L}.
Here we briefly repeat the mathematical description of the model. 
The Fokker-Planck equation that governs electron acceleration, in general, can be written as
 \beqa	
     {\partial f_{\rm ac}\over\partial t} & = &
     {\partial \over \partial E} \left[D(E) {\partial f_{\rm ac} \over \partial E} \right]
   + {\partial\over \partial E} \{[ A(E) - {\dot E}_L ] f_{\rm ac}\} 
 \nonumber \\
 & &  - {f_{\rm ac}\over T_{\rm esc}(E)} + \dot{Q}(E) \,,
 \label{FKeq_code} \eeqa
where $f_{\rm ac} \equiv f_{\rm ac}(t,E)$, in units of electrons $\pcmc \pkeV$,
is the angle-integrated and spatially averaged electron distribution function in energy space
(subscript ``ac" denotes the acceleration region), 	
$E$ is the electron kinetic energy,
$D(E)$ is the energy diffusion coefficient, $A(E)$ the direct acceleration rate, 
$T_{\rm esc}(E)$ the particle escape time, $\dot{Q}(E)$ is the rate of background electrons
being supplied to the acceleration region that serves as a source term, and finally
${\dot E}_L = {\dot E}_{\rm Coul} + {\dot E}_{\rm synch}$	 
is the absolute value of the net energy loss rate that is a sum of the Coulomb and synchrotron loss rates.

In order to solve equation (\ref{FKeq_code})
one must first evaluate all the terms. The energy loss rates ${\dot E}_{\rm Coul}$ and ${\dot E}_{\rm synch}$ 
are well known (see eq.~[18] of PL04) and the source term $\dot{Q}(E)$ 
is to be prescribed with a specific model (assumed to be a thermal or Maxwellian distribution here). 
The central task is then to obtain $D(E)$, $A(E)$, and $T_{\rm esc}(E)$.
They are determined through the dispersion relation of the plasma waves (PL04 eq.~[28]) 
and the wave-particle resonance condition (PL04 eq.~[4]). Following PL04, we assume a fully ionized H
and $^4$He plasma with a relative abundance of electron/proton/$\alpha$-particle=1/0.84/0.08,
and a broken power-law spectrum of the turbulence given by equation~(29) in PL04 with the relevant
spectral indices $q_l=2$, $q=-1.7$ (Kolmogorov value), and $q_h=-4$. The characteristic acceleration rate
$\tau_p^{-1}$ given by equation~(30) in PL04 represents the rate of wave-particle interaction
and depends on the level of turbulence.		



Once all the coefficients have been evaluated, equation (\ref{FKeq_code}) is solved numerically 
using the flux conservative finite difference scheme of \citet{Chang.Cooper1970JCoPh...6....1C}
described in \citet{ParkB1996ApJS..103..255P}. 
Here we assume a homogeneous acceleration region and obtain a steady state solution
of $f_{\rm ac}(E)$ (i.e., $\partial / \partial t=0$).
The angle-integrated flux in the acceleration region is		
 $F_{\rm ac}(E) = v_e f_{\rm ac}(E)$,
where $v_e$ is the electron velocity. We then calculate the flux of electrons that escape the 
acceleration region and enter the transport region of the flare loop,
 \beq F_{\rm esc}(E) = {f_{\rm ac}(E) \over T_{\rm esc}(E)} L \,.
 \label{FescEq} \eeq

\subsection{Particle Transport}
\label{subsect_trans}

The flux $F_{\rm esc}(E)$ is then input to the particle transport code 
\citep{LeachJ1981ApJ...251..781L, McTiernanJ1990ApJ...359..524M} 
which calculates the electron distribution in energy and pitch-angle space
and its variation with distance while the electrons spiral down 
magnetic field lines into deeper layers of the atmosphere. 
The code numerically solves the fully relativistic, steady-state, Fokker-Planck equation
(i.e., eq.~[1] in \citealt{McTiernanJ1990ApJ...359..524M}), 
which is similar to equation (\ref{FKeq_code}) here and includes energy loss (no energy diffusion)
and pitch angle diffusion due to Coulomb collision, 
and pitch angle changes due to magnetic mirroring and synchrotron radiation.
Following \citet{McTiernanJ1989PhDT.142M}, we neglect return currents 
\citep{SyniavskiiD1994ApJS...90..729S,ZharkovaV1995A&A...304..284Z}.

The variable%
 \footnote{In practice, the numerical code equivalently solves for 
 $F(E,\mu,s)/\beta^2 \equiv c\Phi a(s)/a_0$, 		
 where $\Phi$ is defined in \citet{McTiernanJ1990ApJ...359..524M} and $\beta= v_e/ c$.}
to be solved in the transport equation is the electron flux spectrum $F(E,\mu,s)$ as a function of energy $E$, 
cosine $\mu=\cos \theta$ of pitch angle $\theta$, 
and distance $s$ from the injection site at the boundary of the acceleration region.
 $F(E,\mu,s) d\mu$ has units of electrons $\ps \, \pcms \, \pkeV$ 		
and is evaluated as
 \beq F(E,\mu,s) = v_e f(E,\mu,s) {a(s) \over a_0} \,,		
 \label{Fluxeq} \eeq 
where $f(E,\mu,s)d\mu$ is the number density distribution function in units of
electrons~$\pcmc \, \pkeV$			
(cf., the angle-integrated number density $f_{\rm ac}(E)$ in the above acceleration code),
and we integrate the differential electron flux $v_e f(E,\mu,s)$ over the cross-sectional 
area $a(s)$ of the loop and then normalize it by a constant equivalent area $a_0$. 	
As noted earlier, here we assume a constant $a(s)=a_0$ for simplicity, which means
a uniform magnetic field along the loop and thus no magnetic mirroring.

In addition to the injected electron flux $F_{\rm esc}(E)$ from the acceleration code,
the transport code requires the knowledge of the ambient density and abundance distribution along the loop. 
(1) 
Here we assume that $F_{\rm esc}$ is isotropic in pitch angle, representing the consequence of frequent scatterings
of electrons by turbulence in the acceleration region. This assumption is consistent with
the nearly isotropic, rather than beamed, distributions inferred from 
center-to-limb variations of HXR and $\gamma$-ray fluxes and spectral indices in observations
obtained by the {\it Solar Maximum Mission} \citep{McTiernanJ.c2limb.1991ApJ...379..381M}, 
and more recently from atmospheric albedo due to Compton back-scattering
in \hsi flares \citep{Kontar.Stereoscopy.2006ApJ...653L.149K, Kasparova.albedo.2007A&A...466..705K}.
The injected flux at the top ($s=0$) of each leg of the loop is then
$F(E,\mu,s) |_{s=0} = F_{\rm esc}(E) / \int_{-1}^1 d\mu $
which is equivalent to a uniform distribution in the $4 \pi$ solid angle 
integrated over the $2 \pi$ range of the azimuthal angle $\phi$ assuming axis-symmetry.
With the symmetric assumption, the steady state calculation is performed in only one leg of the loop.
We impose a symmetric (or reflective) boundary condition at $s=0$,
where a particle leaving the computational domain is reflected back with identical energy 
but opposite pitch angle cosine, mimicking a particle coming from the other leg of the loop.
(2) As to the background atmosphere, we assume a fully ionized hydrogen plasma
whose distribution is taken from the result of the HD code described next.	


\subsection{Hydrodynamics}
\label{subsect_HDmodel}

Hydrodynamics in the transport region is calculated with the NRL solar flux tube code (MEL89)
based on \citet{MariskaJ1982ApJ...255..783M}. 
The code assumes a two-fluid plasma composed of electrons and ions
that can only move along the magnetic field in a flux tube.	
The user-specified geometry of the tube (a uniform quarter-circle in our case)
is characterized by the tube cross-sectional area $a(s)$ and the component of the gravitational acceleration 
along the tube. The code solves the time-dependent, one-dimensional equations of mass, momentum, and 
energy conservation (see eqs.[1]--[3] in MEL89). The independent variables are the 
mass density $\rho(s)$, fluid velocity $v(s)$, and temperature $T(s)$ which we assume to be the
same for electrons and ions. Because of small masses of electrons, we neglect the momentum loss
of energetic electrons to the background plasma.

The volumetric heating rate in the energy equation is
 \beq S(s)= S_e(s) + S_0 \,, \label{Seq}\eeq
where $S_e(s)$ represents heating by energetic electrons, which is provided by the transport code 
(see \S\ref{sect_comb-codes}), and $S_0=8.31\E{-3} \ergs \ps\pcmc$ (MEL89) represents uniform background heating, 
presumably caused by coronal heating in the quiet sun active region.
The conductive flux $F_{\rm cond}$ and heating rate $S_{\rm cond}$ are
 \beq F_{\rm cond} = - \kappa \frac{\partial T}{\partial s}, \,\,\,\,\,\,
 S_{\rm cond} =  - \frac{\partial F_{\rm cond}}{\partial s} \,,
 \label{FScondEq} \eeq
where $\kappa$ is the thermal conductivity. The radiative energy loss rate is
 \beq L_{\rm rad}= n_e n_p \Phi(T) \,,
 \label{radlossEq} \eeq
where $n_e$ and $n_p$ are the electron and proton number density, respectively, which are equal
by our assumption of fully ionized hydrogen plasma, and $\Phi(T)$ is the 
optically-thin radiative loss function (MEL89) which has its maximum at $T \simeq 1-3\E{5} \K$. 

We select an adaptive mesh of 450 grids that move with time to optimize spatial resolution in the dense chromosphere
and near sharp jumps at the TR and evaporation front. 
This mesh is also shared by the transport and bremsstrahlung radiation codes in our new model.
A reflective (or symmetric) boundary condition is imposed at both the
upper (loop apex) and lower (deep in the chromosphere) boundaries of the
transport region (see Fig.~\ref{model-geometry.eps}), such that the system remains closed.

\subsection{Bremsstrahlung Radiation}
\label{subsect_brem}

Having obtained the electron flux from the transport code and the background density
from the HD code, we calculate the {\it thin-target} bremsstrahlung radiation intensity
or photon emission rate, $I(\e, s)$, as a function of photon energy $\e$ and distance $s$.	
$I(\e, s)$ (photons~$\ps \, \pcmc \, \pkeV$) is defined as 	
 \beq  
  I(\e, s) = \int_\e^\infty dE  \left[ n_p(s) {d \sigma(\e,E) \over d \e} \right] F_{\rm int}  \,,
 \label{nonthBremEq} \eeq
where $d\sigma(\e,E)/d\e$ is the angle-averaged
differential bremsstrahlung cross-section given by \citet{KochH1959RvMP...31..920K},  
and $F_{\rm int}=\int_{-1}^1 F(E, s, \mu) d \mu$ is the angle-integrated electron flux.
Substituting $F_{\rm int}$ with the acceleration region flux $F_{\rm ac}$
gives the LT emission $I_{\rm LT}(\e)$,
while identifying $F_{\rm int}= L F_{\rm thick}$ yields the spatially integrated {\it thick-target} emission 
$I_{\rm thick}(\e)$ \citep{BrownJ1971SoPh, PetrosianV1973ApJ...186..291P}. 
Here $F_{\rm thick}=v_e f_{\rm thick}$ is the {\it equivalent} thick-target electron flux, 
given by the corresponding number density (\citealt{PetrosianV1999ApJ...527..945P}; PL04)
 \beq
 f_{\rm thick}(E) = {1 \over L \dot{E}_{L}} \int_E^\infty F_{\rm esc}(E') d E' \,.		
 \label{fthickeq} \eeq 
The equivalent FP emission is the spatially averaged photons emitted
below the TR (located at $s_{\rm tr}$),	 	
$I_{\rm FP}(\e) = \int_{s_{\rm tr}}^{s_{\rm max}} I(\e, s) ds / (s_{\rm max}-s_{\rm tr})$,	
where $s_{\rm max}$ is the distance at the lower boundary of the loop.
If the coronal is negligibly tenuous and the column depth at $s_{\rm max}$ 	
is large enough to stop all HXR producing electrons of interest, 
$(s_{\rm max}-s_{\rm tr}) I_{\rm FP}(\e)$ approaches $I_{\rm thick}(\e)$.

%
 \begin{figure}[thb]        
 \epsscale{1} 
 \plotone{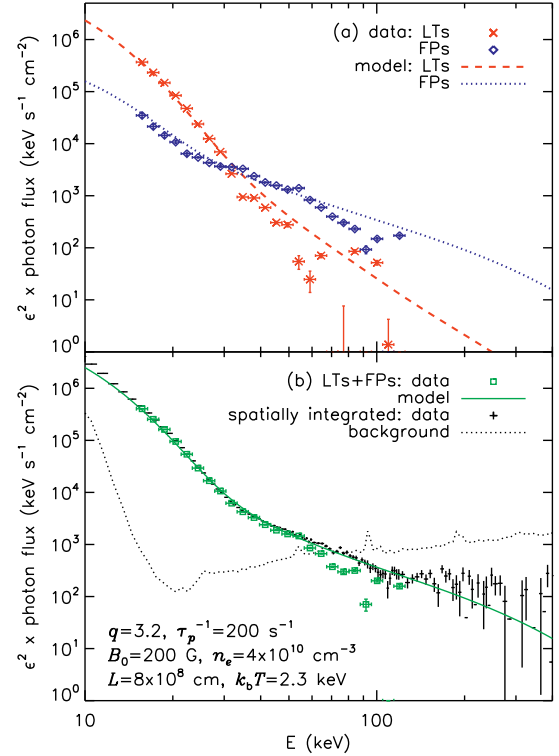}    
 \caption[Spectral fits of the 2002 August 03 X1.0 flare]
 {Photon fluxes multiplied by energy squared ($\e^2$) during the main HXR peak	
 of the 2002 August 3 X1.0 flare and spectral fits using the stochastic acceleration model of PL04. 
   ({\it a})~Summed fluxes of all LT sources	
 and all FP sources 	
 \citep[see Fig.~2.6 of][]{LiuW2006PhDT........35L}.	
   ({\it b})~Sum ({\it squares}) of the LT and FP fluxes shown in ({\it a}) and sum ({\it solid line})
 of the corresponding model fits. Overplotted are the spatially integrated spectrum ({\it plus signs})
 and the corresponding preflare background ({\it dotted line}).
 The legend lists parameters used in the model. (We thank Siming Liu and Yan-Wei Jiang for help
 in producing this figure.)
 }
 \label{spec_2080327.eps}
 \end{figure}
Comparison between the HXRs modeled here and those observed by \hsi satellites
can serve as a unique diagnostic tool and will be pursued in a future publication. Here we show
an example (Fig.~\ref{spec_2080327.eps}) of how well observed LT and FP fluxes can be fitted 
with the above equations
using a spectrum of accelerated electrons given by the PL04 model.

\section{Combining Particle and Hydrodynamic Codes}
\label{sect_comb-codes}

The main task in this study is to combine the Stanford particle code and the NRL HD code.
Here we assume that particle acceleration acts as an independent 
driver of the simulation and is not affected by the temperature or density evolution
in the transport region of the loop.	
The task is thus reduced to make the transport module of the particle code
and the HD code communicate interactively in real time. 

\subsection{Electron Heating Rate}
\label{subsect_heatingRate}

The rate of collisional heating to the background plasma, $S_e$ ($\ergs \, \ps \, \pcmc$), 		
equals the rate of energy loss from the energetic electrons. This can be calculated from the electron 
distribution given by the transport code in two equivalent ways, the second of which is used
in this study.

(1) $S_e$ can be evaluated from the energy loss rate ${\dot E}_{\rm Coul}$ due to Coulomb collisions as
 \beq  
  S_e(s) = \int_{E_{\rm min}}^{E_{\rm max}} dE \int_{-1}^1 f(E,\mu,s) {\dot E}_{\rm Coul} d\mu \,,
 \label{fdepCoulEq} \eeq
where $[E_{\rm min}, E_{\rm max}]$ is the range of the energy bins used in the simulation, and
the electron distribution function $f(E,\mu,s)$ can be obtained from the corresponding electron 
flux $F(E,\mu,s)$ via equation (\ref{Fluxeq}).

(2) Alternatively, one can calculate the net downward energy flux carried by the electrons,
 \beq  F_{\rm erg}(s) = {a_0 \over a(s)} \int_{E_{\rm min}}^{E_{\rm max}} dE 
       \int_{-1}^1  \mu E F(E, s, \mu) d\mu \,,
 \label{ErgflxEq} \eeq
and differentiate it to obtain the net energy gain to the background plasma in a unit volume,
 \beq  S_e(s) = - d F_{\rm erg}(s) /ds \,,
 \label{fdepErgflxEq} \eeq
where $\mu E F(E, s, \mu)$ is the energy flux 	
along the loop	
and the factor $a_0/a(s)$ accounts for the variation of the loop cross-sectional area. 
This approach is practically equivalent to equation (\ref{fdepCoulEq}), because 
in the HXR energy range the combination of 
synchrotron and bremsstrahlung radiation only constitutes a negligible fraction 
($\lesssim 10^{-4}$) of the energy loss of a fast electron due to Coulomb collisions.%


\subsection{Code Communication}
\label{subsect_code-commu}

It is desirable that the particle and HD codes communicate at each time step during 
the time advance. The current transport code, however, can only provide a steady-state solution
and does not have a time-dependent capability. This can be remedied by carefully selecting
the communication time interval $\Delta t$, because particle transport occurs on a much shorter timescale
than hydrodynamics. This interval should be as short as possible 	
provided that a steady-state transport solution can be reached. By considering the electron 
``lifetime" (see eq.~[9] of \citeauthor{PetrosianV1973ApJ...186..291P} 1973), 
which is determined by the energy loss time in a given loop geometry and 
atmospheric density distribution, \citet[][his \S 7.2.4]{LiuW2006PhDT........35L}
found the optimal interval to be $\Delta t= 2 \s$.

The remaining question is what heating rate the HD code should use during its time advance 
between adjacent communications with the particle code.
Let us first change the independent variable of $S_e$ from distance $s$ to column depth 
$N(s)= \int_0^s n_e(s') ds'$,
 \beq
  S_e[N(s)] = S_e(s) / n_e(s) \,,
 \label{Seqs2N} \eeq
noting $S_e(s) ds = S_e(N) dN$ and $dN= n_e ds$. Here $S_e(N)$ is in units of $\ergs \, \ps$.
We have assumed a loop of uniform cross section and thus no magnetic mirroring,
and here we further neglect synchrotron loss ${\dot E}_{\rm synch}$ (valid for $\lesssim$1~MeV electrons).
Under these assumptions, the electron flux $F[E, \mu, s(N)] \equiv F[E, \mu, N(s)]$ 
is a function of column depth $N$, independent of distance $s$, and so does the heating rate 
$S_e(N)$ calculated from $F(E, \mu, N)$.

The HD response timescale is characterized by sound travel time, which is 84~s
for a sound speed of $166 \, \km \ps$ at $T=10^6 \K$ in a $14 \Mm$ long loop here.		
Since $\Delta t= 2 \s \ll 84 \s$, we assume that $S_e(N)$ is constant in time 
between adjacent code communications. During this $\Delta t$,
the spatial distribution of the heating rate $S_e(s,t)$ varies with time merely
according to the redistribution of density and thus the variation of column depth, 
 \beq
  S_e(s,t) = S_e(N) n_e(s, t) \,.
 \label{SeqN2s} \eeq
In practice, at a given time $t$ and distance $s$, we first identify its
column depth $N(s, t)$, which is then used to evaluate the heating rate $S_e(N)$, 
and then we apply 	
the local density to obtain $S_e(s, t)$ by equation~(\ref{SeqN2s}).

\begin{figure}[thb]	      
 {\footnotesize
 \setlength{\unitlength}{0.55cm}	
 \begin{picture}(16, 8)

  \put(10,6.5){\framebox(4,1)[c]{Start}}
  \put(12,6.5){\vector(0,-1){1}}

  \put(10,4.5){\framebox(4,1)[c]{HD initial state}}
  \put(10,5){\vector(-1,0){5}}  \put(6,5.2){Density $n_e(s)$}

  \put(1,4.5){\framebox(4,1)[c]{Particle 1st run}}
  \put(3,4.5){\vector(0,-1){1.5}}
  \put(3,3){\vector(1,0){9}}        \put(5,3.2){Heating rate $S_e(N)$}
  \put(12,3){\vector(0,-1){1.5}}
 
  \put(10,0.5){\framebox(4,1)[c]{HD run $\Delta t$=2~s}}
  \put(10,1){\vector(-1,0){5}}  \put(6,1.2){Density $n_e(s)$}

  \put(1,0.5){\framebox(4,1)[c]{Particle run}}
  \put(3.5,1.5){\vector(0,1){1.5}}
 \end{picture}
 } 
 \caption[]{Task flow chart for communications between the particle and HD codes.}
 \label{fig_flowchart} \end{figure}
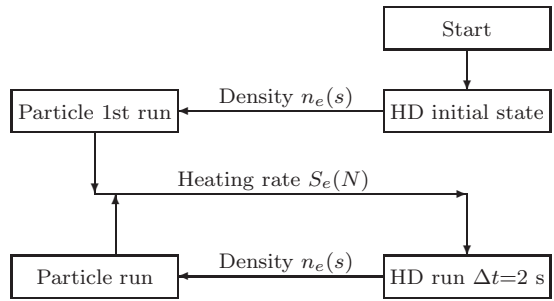

Communications between the two codes are summarized in Figure~\ref{fig_flowchart}.
First, the HD code passes the initial
density distribution to the particle code, which then runs its first steady-state calculation and returns the heating 
rate $S_e(N)$ as a function of column depth to the HD code. Next the HD code repeatedly
converts $S_e(N)$ to $S_e(s, t)$ as a function of distance at each time step using the latest 
density profile. Once the HD code advances a time interval of $\Delta t = 2 \s$, 
it passes the updated density distribution back to the particle code, 
which starts the next cycle of iteration.
%
%

\section{Simulation Runs}
\label{sect_result}


%
\begin{table*}[th]    
\footnotesize	
\caption[]
{Summary of simulation runs.}
\tabcolsep 0.04in
\begin{tabular}{lllllllllll}		
\tableline\tableline
Runs & \multicolumn{2}{c}{Injected Electron}  & Particle   & $v_{\rm max}$   & $t_{v_{\rm max}}$ & $v_{\rm min}$  & $t_{v>100}$ & $t_{\rm apex}$ & $T_{\rm max}$   & $n_{e,\rm max}$    \\
       \cline{2-3}  
     & Spectrum   &  Ang. Distr.  & Transport & ($\kmps$) & (s)           & ($\kmps$)& (s)         & (s)        & ($10^6 \K$) & ($10^{10} \pcmc$)  \\
\tableline
O(ld)       & power law     & beamed & analy. approx.   &  565 & 35 & $-115$ & 10& 29 & 21.1 & 6.96 \\
H(ybrid)    & power law     & beamed & Fokker-Planck    &  570 & 35 & $-106$ & 9 & 28 & 22.0 & 7.44 \\
N(ew)       & stoch. accel. & isotropic & Fokker-Planck  &  598 & 36 & $-102$ & 9 & 23 & 25.9 & 8.44 \\
\tableline
\end{tabular}

{\bf Notes.} 
Runs O and H have an identical injected power-law electron flux, with a spectral index $\delta=6$ and
low-energy cutoff $E_c=15 \keV$;
$v_{\rm max}$ and $t_{v_{\rm max}}$: maximum (upflow, $v>0$) velocity and its time stamp; 
$v_{\rm min}$: minimum (downflow, $v<0$) velocity (in the upper chromosphere);
$t_{v>100}$: time when the upflow velocity exceeds $100 \kmps$;
$t_{\rm apex}$: time when the evaporation front (density jump) reaches the loop apex;
$T_{\rm max}$ and $n_{e,\rm max}$: maximum coronal temperature and electron density.
All runs have the same peak energy deposition (electron heating) flux for the loop as a whole, 
$F_{\rm max}= 2.67 \E{10} \ergs \s^{-1} \pcms$.

\label{table_cases} 
\end{table*} 
%
We have performed three simulation runs (see Table~\ref{table_cases})
to test the relative effects of different processes.		
 (1) In the first run, which we refer to as the Old Model (abbreviated by ``O"),
we assumed an injection of electrons of a power law at the LT, and evaluated the heating
rate and HD response as in the MEL89 model. This model does not calculate particle transport properly. 
 (2) In the second simulation, we still injected power law electrons and evaluated
electron transport and heating along the loop using our transport code. We call this 
the Hybrid Model (or ``H").
 (3) Finally, we employed our most realistic model, where we evaluated from our acceleration code (PL04)
the spectra of electrons at the LT acceleration site and those escaping the LT region.
We also calculated transport and heating using our transport code. We call it the
New Model (or ``N").
 In all cases, we assumed an identical initial HD state as shown in Figure~\ref{model-geometry.eps},
and calculated the HD response using the MEL89 code. We assumed the dynamic or modulation profile
of the number of injected electrons (power law for Runs~O and H and $\dot{Q}$ for Run~N)
to be a triangular shape with a rise and fall to be 30~s each. Beyond this first 60~s
of the impulsive phase, while the electron heating rate $S_e$ was set to zero, 
we continued the computation into the decay phase until $t=90 \s$.
 \begin{figure}[thbp]      
 \epsscale{1} 
 \plotone{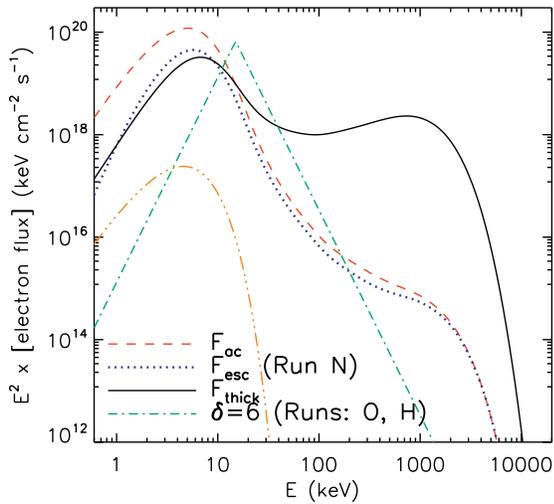}   
 \caption[]
 {Electron flux spectra $F(E)$ times $E^2$ at the peak time ($t=30$~s):
 (1) acceleration region flux $F_{\rm ac}$,
 escaping flux $F_{\rm esc}$, and equivalent thick-target flux $F_{\rm thick}$ for Run~N, 
 and (2) injected flux for Runs~O and H ($\delta=6$ above $E_c=15$~keV). The triple-dot-dashed line
 (arbitrary scale) represents the source term $\dot{Q}(E)$ of a Maxwellian (thermal, $k_b T=1.53 \keV$) 
 distribution used in Run~N,  which peaks at $E=3k_b T$		
 in this $E^2F(E)$ plot. 	
 } \label{init_elspec.eps}
 \end{figure}
%

\subsection{Run~O: Old Model}	
\label{subsect_RunO}

We first computed the HD response using the same heating function 
and almost identical control parameters as the ``Reference Calculation" of MEL89. 
This model is based on the analytical solution of \citet{Emslie1978ApJ...224..241E}
that includes only energy loss and pitch angle {\it growth} of injected electrons.
By the MEL89 assumption, the injected electron flux is downward beamed ($\mu_0=1$),
and its energy spectrum (see Fig.~\ref{init_elspec.eps}) is a broken power law $\propto E^{-\delta}$ with
an index $\delta=6$ above and $\delta=-2$ below the cutoff (``knee") energy $E_c= 15$~keV. 
Here the differences are: (1) our peak energy flux,
i.e., parameter $F$ in equation~(9) of MEL89, is $2.67 \E{10} \ergs \pcms \ps$, 
while they used $5 \E{10} \ergs \pcms \ps$, and (2) we assume a fully ionized hydrogen plasma
while they included helium of 6.3\% of hydrogen in number density. The latter difference only 
changes the {\it absolute} mass density by 11\%, while the {\it relative} differences between
our models may not be affected. 	

\subsection{Run~H: Hybrid Model}
\label{subsect_RunH}

Here we used our Fokker-Planck transport code in place of the
approximate analytical expression used above to evaluate the heating rate.
We injected electrons of an identical power-law spectrum with a narrow-Gaussian 
($\sigma_{\mu}=0.01$) pitch-angle distribution to emulate the beamed distribution in Run~O. 
The main difference here is that the transport code properly treats 
the {\it diffusion} process of pitch angle change due to Coulomb collision.
For the particle code, the energy space is divided into 200 uniform logarithmic bins
in the range of $511 \times [10^{-3}, 10^3]$~keV, while the pitch angle space
is divided into 100 \citep[vs.~24 used in][]{LiuW2006PhDT........35L} 
uniform bins in the $[0, \pi]$ range.

\subsection{Run~N: New Model}
\label{subsect_RunN}

This is a typical simulation using our new model. It is the same as Run~H,
except that the injected beamed power-law electron flux 	
is replaced with an isotropic flux given by the stochastic acceleration code.
We used the same acceleration parameters as PL04 
(see their Fig.~12), i.e., the characteristic acceleration rate $\tau_p^{-1}=70 \ps$,
$n_e= 1.5 \E{10} \pcmc$, $B_0= 400 \G$, $k_b T = 1.53 \keV$, and the acceleration region size $L = 5 \E8 \cm$.
We modulated the rate ($\dot{Q}$, see eq.~[\ref{FKeq_code}]) of electrons being supplied to the acceleration region
with the same triangular time profile,
such that the peak electron heating		
flux $F_{\rm max}$ equals that of Run~O or H. 

Figure~\ref{init_elspec.eps} shows various electron flux spectra used in this study. In comparison with the
background thermal distribution, both the acceleration region flux $F_{\rm ac}$ 	
and escaping flux $F_{\rm esc}$		
have a quasi-thermal component		
that smoothly extends to a nonthermal tail at high energies. $F_{\rm esc}$ is smaller than $F_{\rm ac}$ 
and their relative difference decreases with energy due to the energy-dependent confinement of electrons
by turbulence in the acceleration region (see eq.~[\ref{FescEq}]). 
Unlike that ({\it dot-dashed}) in Run~O or H, the flux $F_{\rm esc}$ injected into the transport region 
does not invoke any arbitrary low-energy cutoff.	
The two fluxes, however, have similar slopes in the intermediate energy range around 20~keV.
The equivalent thick-target electron flux $F_{\rm thick}$ (see \S\ref{subsect_brem}),	
as expected, has a harder spectrum than $F_{\rm ac}$ and $F_{\rm esc}$	
in the 10--1000~keV range.

\section{Simulation Results: Comparison of Fluid Dynamics}
\label{sect_cmpr}

To determine how much our proper transport and acceleration calculations affect the 
atmospheric response, we compare the results of the three simulations, using Run~H as
the reference case.	

\subsection{Hydrodynamic Evolution}
\label{subsect_HDevol}

 \begin{figure}[thbp]      
 \epsscale{1.24}	
 \plotone{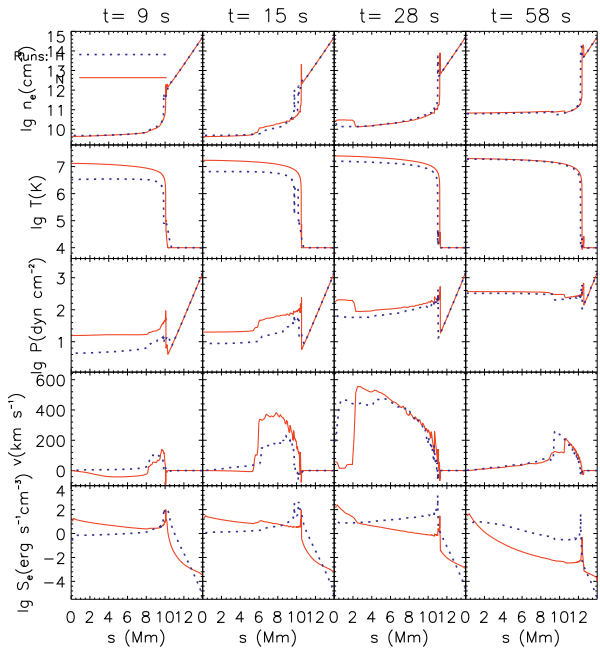}  
 \caption[]
 {Evolution of electron density, temperature, pressure, velocity, and electron heating rate for Runs~H
 and N.		
 } \label{HDmosaic_R.eps}
 \end{figure}
\begin{figure}[thbp]      
 \epsscale{0.8} 
 \plotone{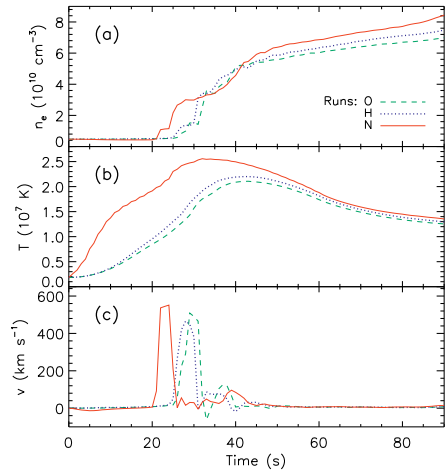}  
 \caption[]
 {History of electron density, temperature	
 and velocity at $s=1$~Mm from the injection site for different simulations.
 } \label{cmpr_history.eps}
 \end{figure}

As shown in Figure~\ref{HDmosaic_R.eps},	
Run~H exhibits similar general HD evolution as described in MEL89 (their Figs.~1 and 2).
Electron heating ($S_e$) is initially concentrated in the upper chromosphere,	
producing overpressure that drives an upflow ($v>0$) and a recoil downflow ($v<0$).
At $t=9 \s$, the upflow velocity exceeds $100 \km \ps$ and a density jump	
or evaporation front has developed slightly above the TR. 
It travels upward and reaches the loop apex at $t=28 \s$. The density jump is then reflected 
back and the material piles up due to the reflective boundary condition imposed, which
can be understood as plasma flow from the other half of the loop in the assumed symmetric geometry.
The upflow reaches its maximum velocity of $v_{\rm max}=570 \km \ps$ at $t=35 \s$, which
is delayed by 5~s from the energy deposition peak at $t=30 \s$. 
Chromospheric evaporation then gradually subsides.	
These features of the temporal evolution can also be seen from the history of various quantities at a fixed
position in the upper corona as shown in Figure~\ref{cmpr_history.eps}. 
Note that, late during the simulation, the coronal temperature gradually decreases mainly due to conductive cooling, 
while the coronal density continues to increase, even after the cease of electron heating at $t=60 \s$.
This is caused by sustained chromospheric evaporation that results from
heating of the chromosphere by the same conductive flux that cools the hot corona.

In comparison, we find that Run~O is very similar to Run~H (see Fig.~\ref{cmpr_history.eps}
and Table~\ref{table_cases}) but less intense. This can be more clearly seen
from various HD quantities and heating and cooling rates during the early phase
shown in Figure~\ref{HDcmpr_multi.eps}. The small differences (on the order of 10\%) 
between the two cases are mainly caused by the slight overestimate of electron heating
of Run~O in the chromosphere (Fig.~\ref{HDcmpr_multi.eps}{\it d}), 
due to its inaccurate way of calculating particle transport noted earlier.
This indicates that, for this specific case, the analytical heating function used in MEL89 provides an acceptable
approximation for the more accurate Fokker-Planck calculation.

In contrast, the HD evolution of Run~N is faster and more intense than that of Run~H
(Figs.~\ref{HDmosaic_R.eps} and \ref{cmpr_history.eps}, Table~\ref{table_cases}), 
despite the same maximum electron heating rate for the two cases.
These differences are primarily due to the different {\bf energy spectra} of injected electrons,
further enhanced by their different {\it pitch-angle distributions}:
%
 \begin{figure*}[thbp]      
 \epsscale{0.323} 	
 \plotone{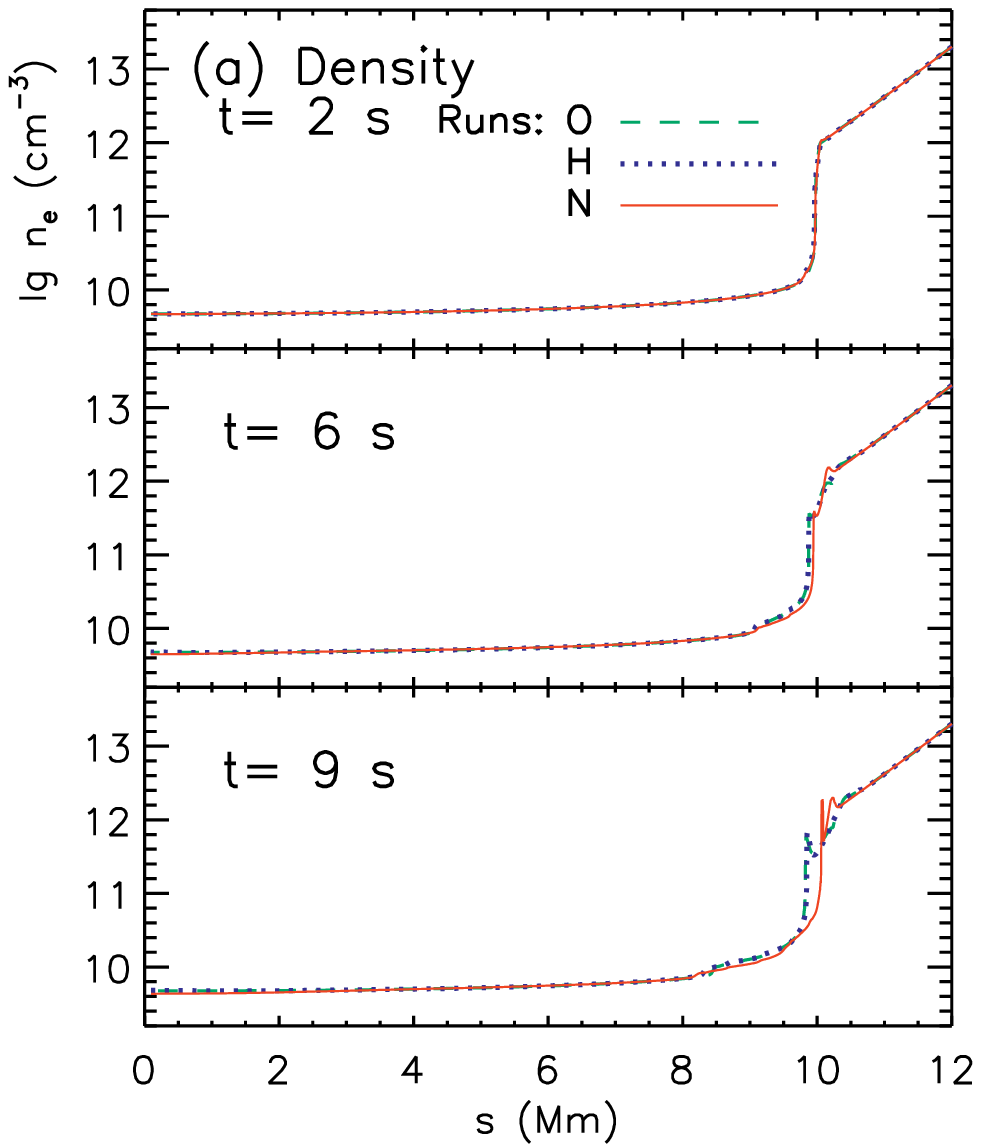} 
 \plotone{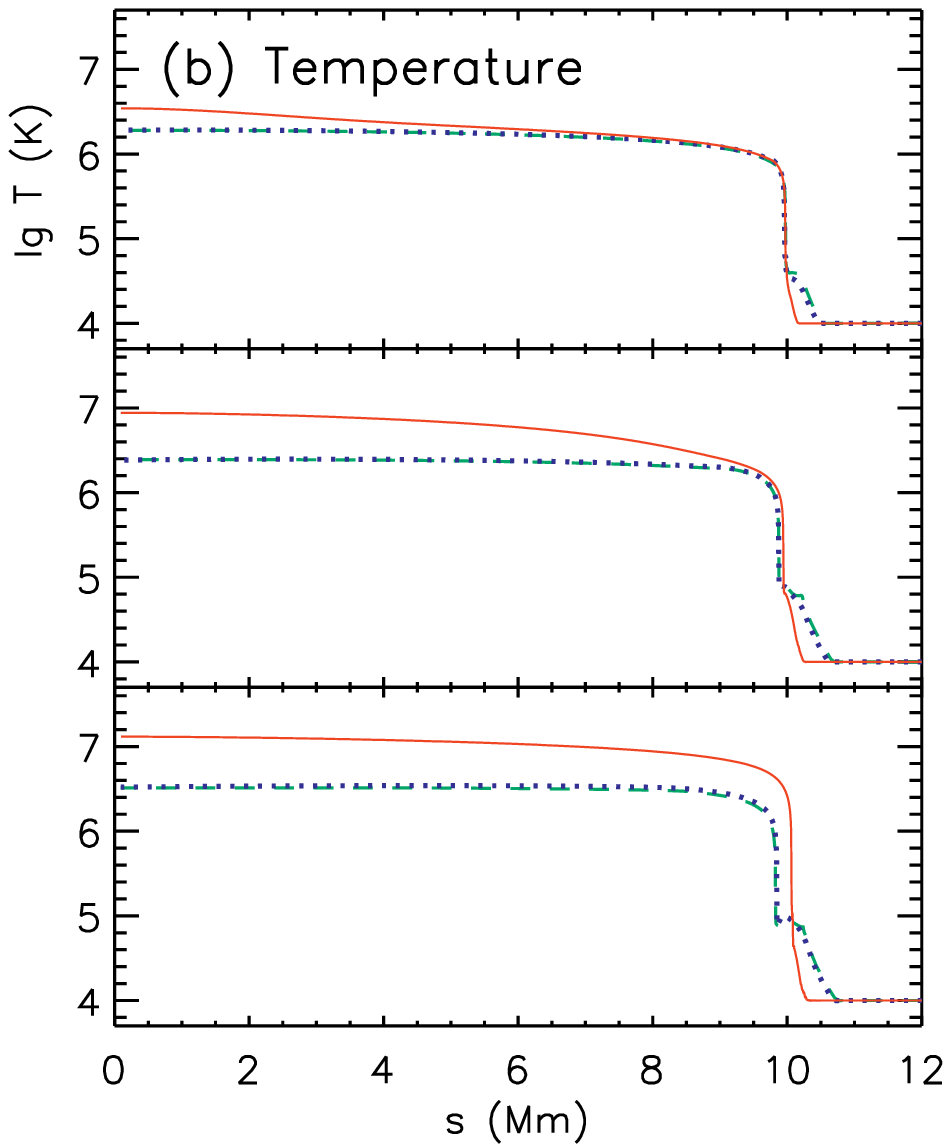} 
 \plotone{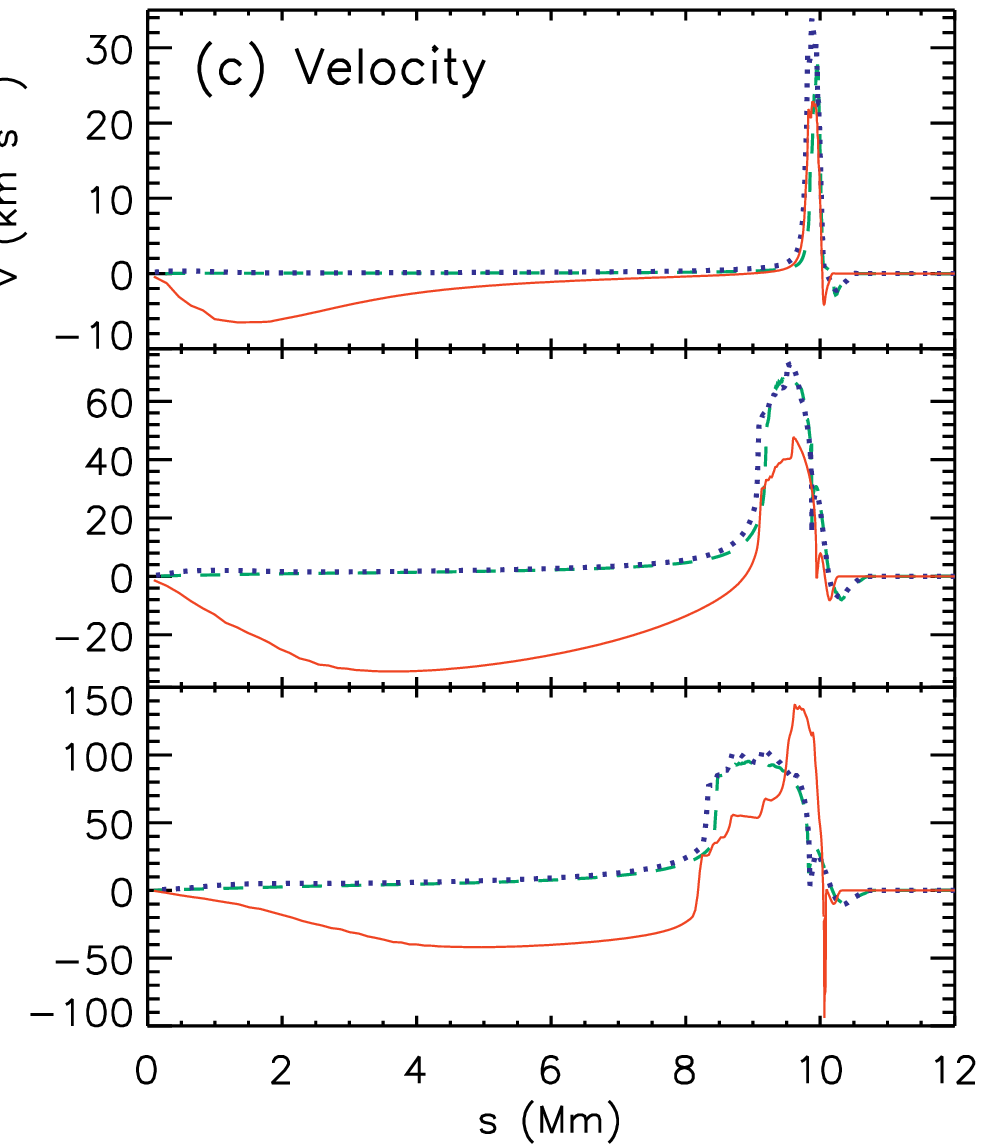} 
 \plotone{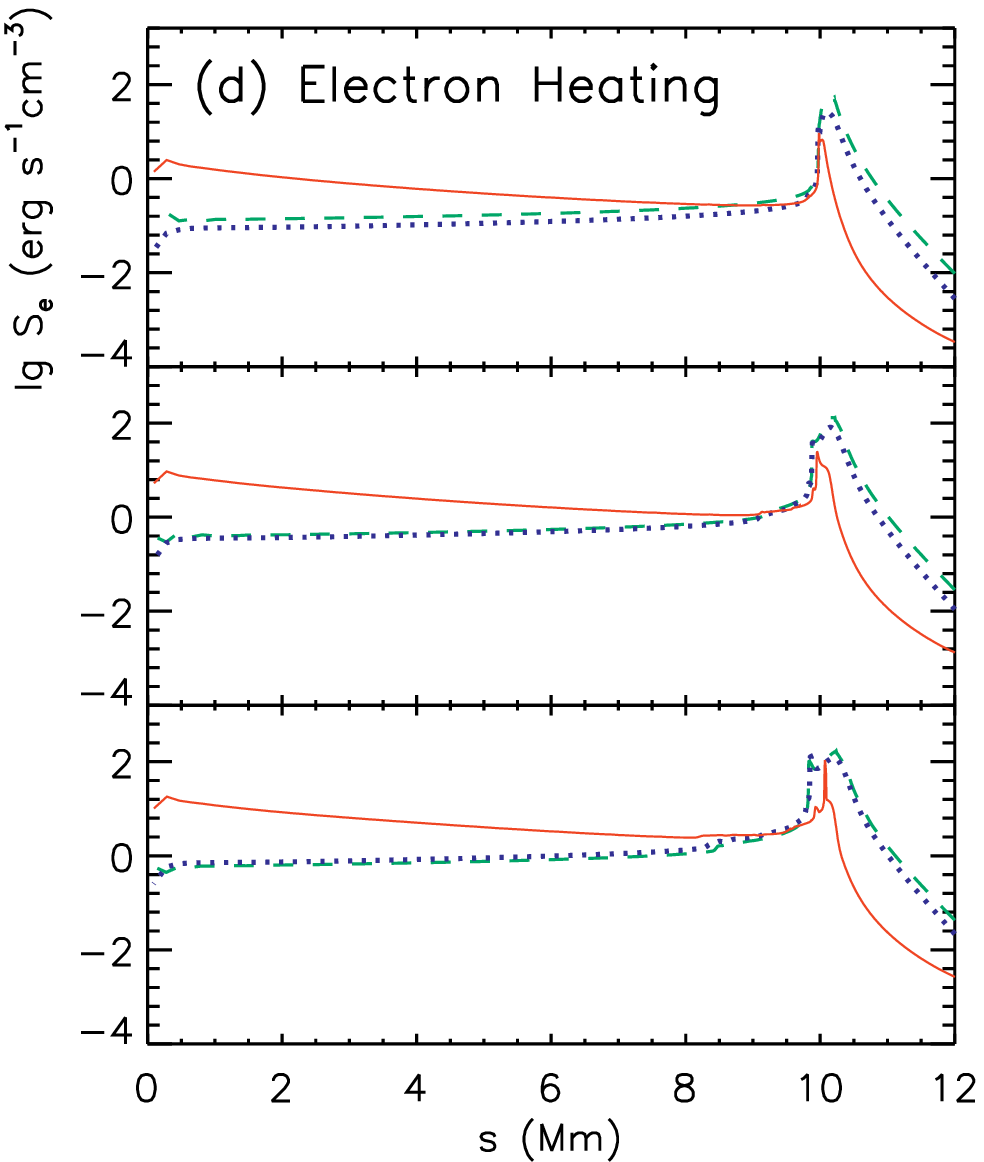} 
 \plotone{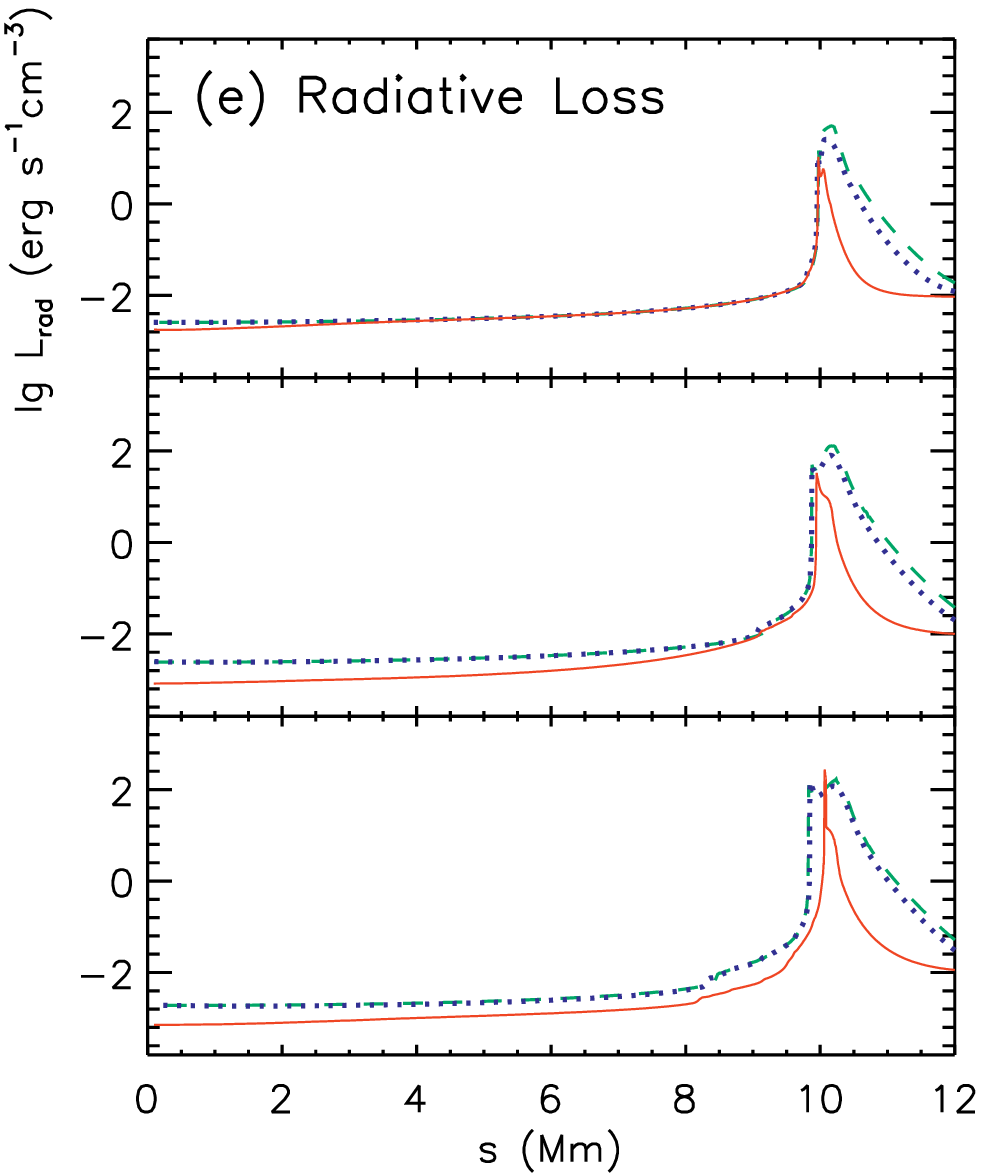} 
 \plotone{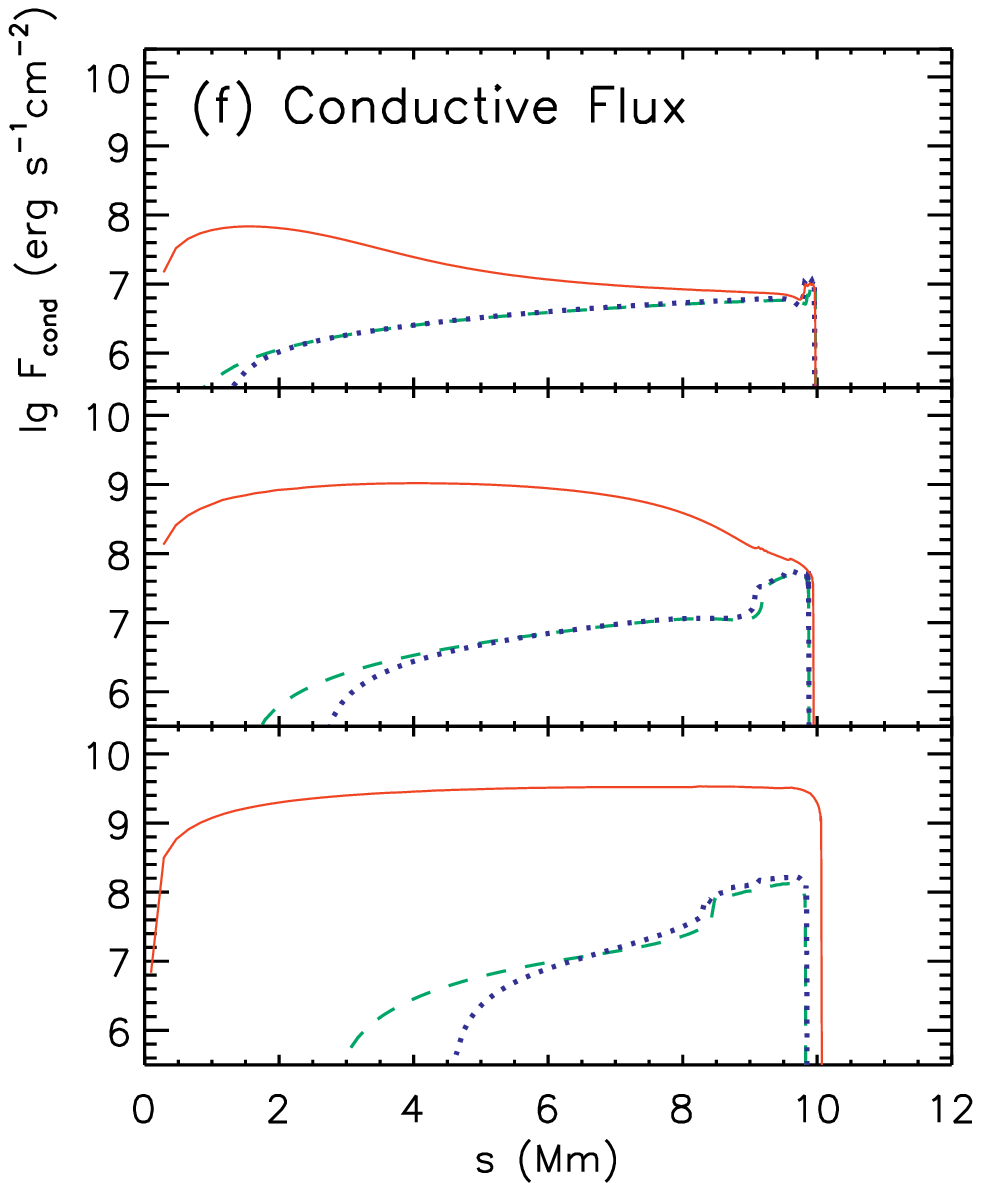} 
 \caption[]
 {Same as Fig.~\ref{cmpr_history.eps} but for snapshots of various quantities as a function
  of distance at selected times early in the flare.
 } \label{HDcmpr_multi.eps}		
 \end{figure*}

 (1) In Run~N, electrons of a few keV in the quasi-thermal component of the spectrum (Fig.~\ref{init_elspec.eps}) 
are small in energy but large in number and thus dominate the total energy content. 
These electrons 	
produce heating at relatively small column depths by Coulomb collisions.
This occurs high in the corona (Fig.~\ref{HDcmpr_multi.eps}{\it d}), where the radiative loss rate $L_{\rm rad}$
(Fig.~\ref{HDcmpr_multi.eps}{\it e}) is relatively small due to the low density and high temperature. 
As a result, significant net heating sets in there, which leads to a local temperature
(Fig.~\ref{HDcmpr_multi.eps}{\it b}) and pressure surge.
This local coronal heating is enhanced by the large effective column depth 
($N_{\rm eff}=N/ \langle \mu \rangle $, where $\langle \mu \rangle $ is the mean pitch angle cosine)
resulting from the isotropic pitch-angle distribution of the injected electrons.
The increased temperature leads to a large downward heat conduction flux (Fig.~\ref{HDcmpr_multi.eps}{\it f}),
and the pressure gradient force drives a downward mass flow in the high corona (Fig.~\ref{HDcmpr_multi.eps}{\it c}).

 (2) In Run~H, contrastingly, the electron spectrum (Fig.~\ref{init_elspec.eps}) peaks at 
the cutoff energy $E_c$, which leads to a deficit in low-energy electrons.
In addition, the pitch-angle distribution here is beamed (rather than isotropic). 
This electron population, on average, penetrates deeper into the atmosphere than
that in Run~N and deposit their energy primarily in the upper chromosphere.
This results in less heating in the corona and stronger and more widespread heating in the chromosphere 
(Figs.~\ref{HDcmpr_multi.eps}{\it b} and \ref{HDcmpr_multi.eps}{\it d}). Consequently, 
in spite of the larger and broader radiative cooling (Fig.~\ref{HDcmpr_multi.eps}{\it e}), the local overpressure
in the chromosphere is stronger than that in Run~N early on,	
which drives a higher velocity upflow (Fig.~\ref{HDcmpr_multi.eps}{\it c}, $t=6\s$).
Also, unlike Run~N, there is no significant
downward coronal heat conduction (Fig.~\ref{HDcmpr_multi.eps}{\it f}) or mass flow.

\subsection{Heating and Cooling}
\label{subsect_heating}

A remaining question is why Run~N has more dramatic overall HD changes in the long run.
To answer this, we examine the relationship between different energy gain and loss terms --- electron heating,
radiative loss, and conductive heating and cooling --- particularly early in the flare and
near the TR where chromospheric evaporation takes place. 

\begin{figure}[thbp]      
 \epsscale{1.2} 
 \plotone{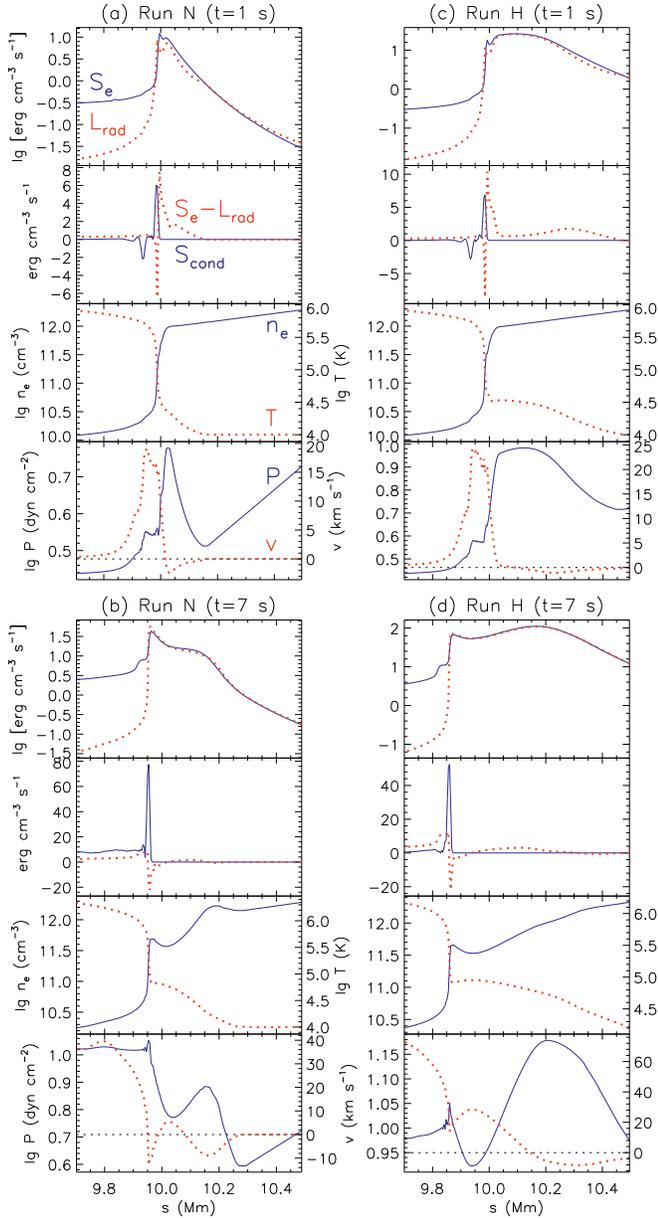}
 \caption[]
 {Comparison between Runs~N ({\it left}) and O ({\it right}) of a detailed
 view near the TR at two selected times. At each time, the quantities in the four
 panels (from top to bottom) are: (1) electron heating rate $S_e$ and radiative loss rate $L_{\rm rad}$,
 (2) $S_e-L_{\rm rad}$ ({\it dotted}) and conductive heating rate $S_{\rm cond}$ ({\it solid}), 
 (3) electron density ({\it left scale}) and temperature ({\it right scale}),
 and (4) pressure ({\it left}) and velocity ({\it right}). 
 Note different vertical scales.	
 } \label{cmpr_heating.eps}
 \end{figure}

(1) In Run~N at $t=1 \s$ (see {\it top panel} of Fig.~\ref{cmpr_heating.eps}{\it a}), the electron heating
rate $S_e$ peaks in the TR because of the sharp increase there in ambient density and
associated collisional energy loss of energetic electrons. So does the radiative loss rate $L_{\rm rad}$,
since it is proportional to $n_e n_p$ and $\Phi(T)$ that peaks at $T\simeq 1-3 \E{5} \K$ which is 
the TR temperature.	
However, due to their different functional dependencies on density
and temperature, $S_e$ peaks at a slightly lower position than $L_{\rm rad}$. Their combination
$S_e-L_{\rm rad}$ ({\it panel 2}, Fig.~\ref{cmpr_heating.eps}{\it a}) thus results in cooling in 
the upper TR 	
and heating in a shallow layer below it in the upper chromosphere. 
Meanwhile, the conductive flux carries energy from the hot upper corona to the upper
TR where localized heating ($S_{\rm cond}$) is produced and counteracts radiative cooling.	
(Conduction is prohibited in the chromosphere where the temperature is maintained at $10^4 \K$.) 
The net energy gain resulting from the interplay of electron heating, radiative cooling, 
and heat conduction is thus localized in the upper chromosphere, where temperature is raised substantially
({\it panel 3}, Fig.~\ref{cmpr_heating.eps}{\it a}). This leads to a local pressure hump
which drives an upflow into the corona and a downflow into the chromosphere
({\it panel 4}, Fig.~\ref{cmpr_heating.eps}{\it a}). 

\begin{figure}[thbp]      
 \epsscale{0.8} 
 \plotone{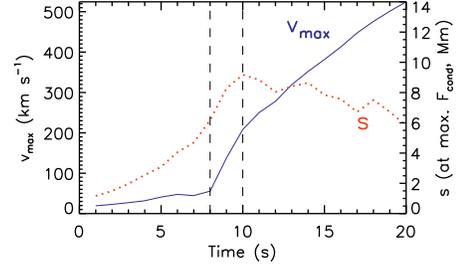}
 \caption[]
 {Early history of the maximum upflow velocity $v_{\rm max}$ ({\it left scale}) and the position 
 of the maximum conductive flux ({\it right scale}) for Run~N. The two vertical lines mark the interval
 when $v_{\rm max}$ experiences a sharp increase while the maximum $F_{\rm Fond}$ rapidly
 approaches the TR.
 } \label{vmax_zccmax.eps}
 \end{figure}

As time proceeds (see Fig.~\ref{cmpr_heating.eps}{\it b}) and chromospheric material is being
heated from $T=10^4 \K$ to $\sim$$10^5 \K$ where $\Phi(T)$ reaches its maximum, radiative loss gradually
overtakes electron heating in the TR and upper chromosphere.
This means that energy directly deposited by electrons in these places is immediately radiated away
and very little is left to heat the plasma. However, as we noted above, a significant portion of
the energy content of the injected electrons in Run~N is deposited in the upper corona
(Fig.~\ref{HDcmpr_multi.eps}{\it d}) where radiative loss
is negligible and then transported by heat conduction to the lower atmosphere. In time, conduction plays 
an increasingly important role in heating the lower corona and TR as
it gradually exceeds the direct net heating or combined electron heating and radiative loss $S_e-L_{\rm rad}$
(Fig.~\ref{cmpr_heating.eps}{\it b}) in these regions. Because of this, as of $t=7 \s$ the location 
of the primary net heating, i.e., the highest local overpressure, and the maximum downflow velocity have shifted 
from the upper chromosphere up to the TR. In fact, the peak of the conductive flux $F_{\rm cond}$
(see Fig.~\ref{HDcmpr_multi.eps}{\it f}) and the region of significant conductive heating
($S_{\rm cond} \propto -d F_{\rm cond} / ds$) below it 	
are initially located in the upper corona and propagate down the loop with time. 
As the $F_{\rm cond}$ peak approaches the TR at $\sim$10~Mm during the interval
of 8--10~s,	
the maximum upflow velocity $v_{\rm max}$ in the loop increases abruptly (see Fig.~\ref{vmax_zccmax.eps}).
This further emphasizes the role played by heat conduction here in redistributing energy deposited by
electrons and in driving chromospheric evaporation.

(2) In Run~H, 	
the injected electron flux with a low-energy cutoff has profound consequences. As noted earlier, 
lack of low-energy electrons makes the chromosphere,
rather than the upper corona, the primary location of direct heating early in the flare
(see Fig.~\ref{HDcmpr_multi.eps}{\it d}).	
Net direct heating $S_e-L_{\rm rad}$ and 
the increase of local temperature and pressure extend from the TR to much deeper layers 
of the chromosphere than in Run~N (Fig.~\ref{cmpr_heating.eps}{\it c}).		
Moreover, since the coronal 	
temperature does not increase rapidly, the downward conductive flux here	
is more than an order of magnitude smaller 	
(Fig.~\ref{HDcmpr_multi.eps}{\it f}). 		
Net direct heating generally dominates over conductive heating when integrating over the volume of 
the lower atmosphere.	
As a result, in comparison with Run~N, a relatively larger portion of the total energy content
of the injected electrons is lost in radiative cooling.
This is why the overall HD
development here is more gentle than that of Run~N, despite the fact that they have the same energy deposition flux.
The primary underlying physics is their different spatial distributions of electron heating $S_e$
caused by their different electron injections.
Note that MEL89 obtained qualitatively similar results when different values of the 
cutoff energy or spectral index were considered.
We note that, later 
during the impulsive phase in Run~H (Fig.~\ref{HDmosaic_R.eps}), 
as the coronal density has increased considerably due to chromospheric evaporation, 
relatively more electron energy is directly dumped in the corona while less in the chromosphere.
Consequently, 	
conductive heating in the TR becomes important.
In this sense, the physical distinction between the two runs gradually diminishes in the late stage	
of the flare.

\subsection{Velocity Distribution}
\label{subsect_vel-distr}

\begin{figure}[thbp]      
 \epsscale{0.57} 
 \plotone{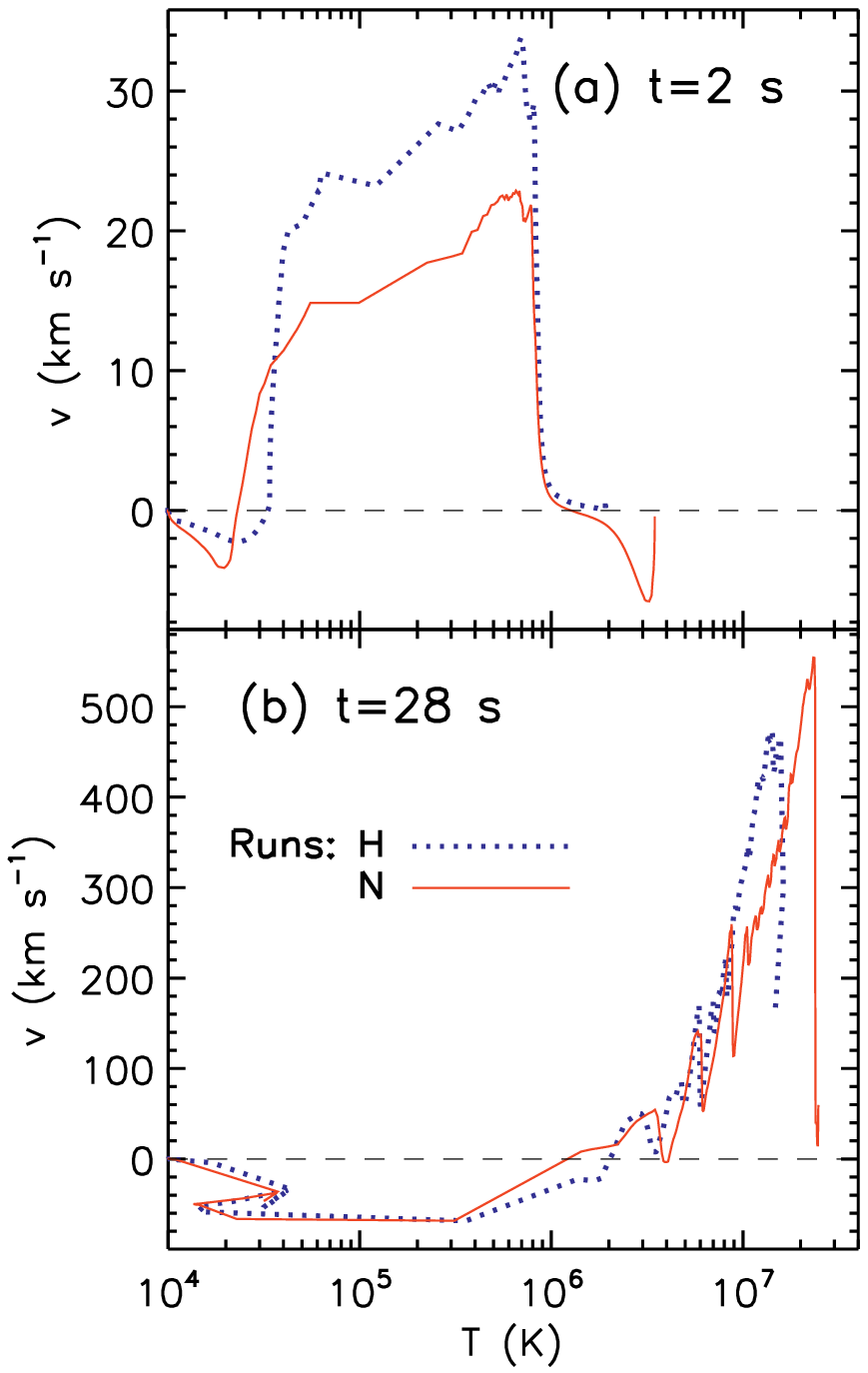}  
 \plotone{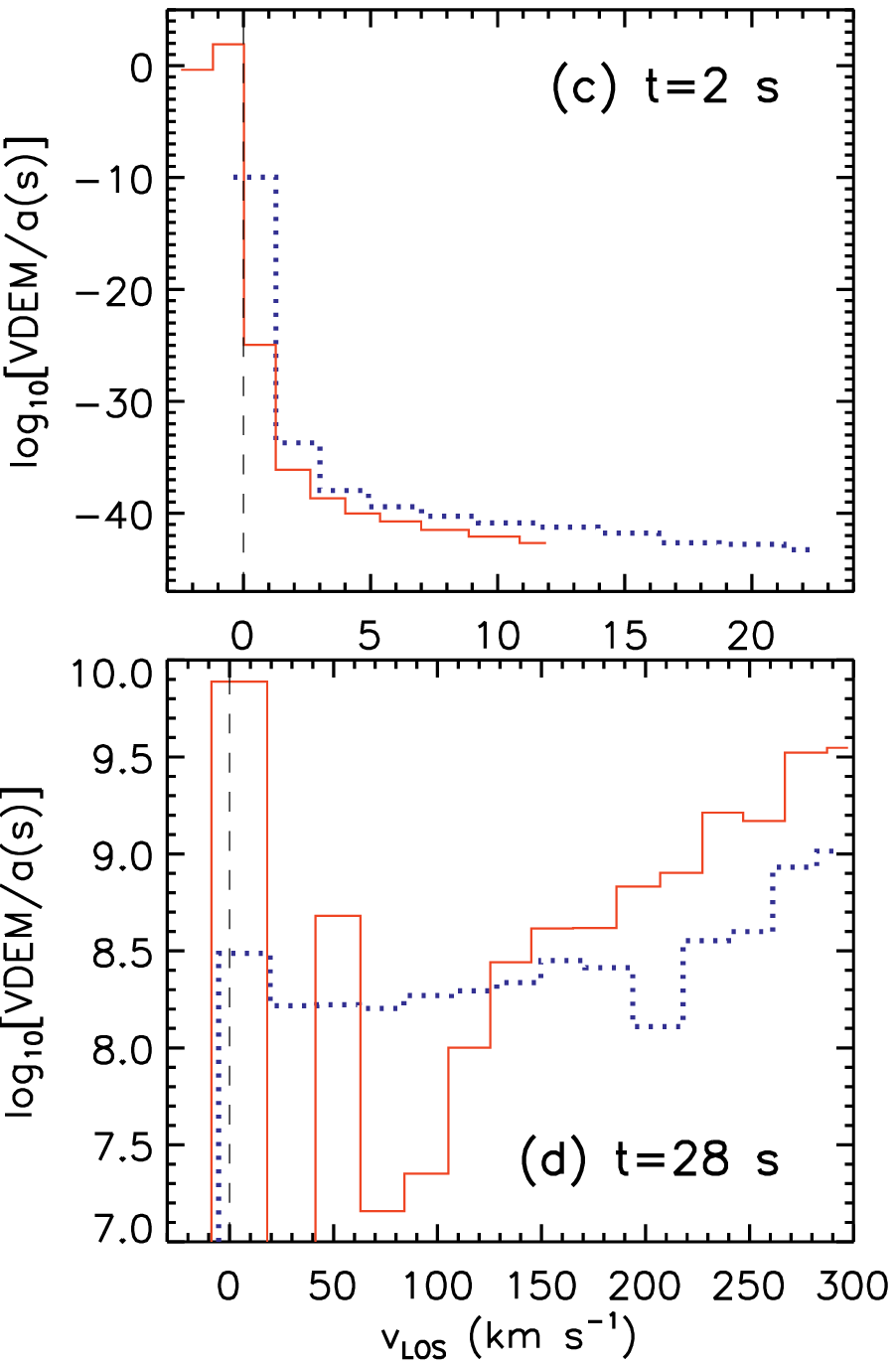}
 \caption[]
 {Comparison between Runs~H and N:		
 velocity vs.~temperature ({\it left}) and corresponding specific velocity differential emission measure 
 (photons $\pcms \, \ps \, {\rm sr}^{-1} \, [\km \, \ps]^{-1}$) vs.~LOS velocity ({\it right}).	
 } \label{cmpr_Tv.eps}	
 \end{figure}

Here we examine observables that can be checked against data:
(1) the temperature dependence of the plasma velocity (Fig.~\ref{cmpr_Tv.eps}, {\it left}),
and (2) the velocity differential emission measure (VDEM; Fig.~\ref{cmpr_Tv.eps}, {\it right}) 
defined by \citet{Newton.VDEM.1995ApJ...447..915N} as the emissivity ($G[T]$) weighted
measure of material with line-of-sight (LOS) velocity $v_{\rm LOS}$.
Assuming that the flare is located at the solar disk center such that the LOS is perpendicular to
the loop apex, we calculated the {\it specific} VDEM/$a(s)$ for the Ca~XIX 3.18~\AA\
line following \citeauthor{Newton.VDEM.1995ApJ...447..915N} 
Here $G(T)$ is given by the Chianti package \citep{YoungP2003ApJS.CHIANTI}.		
%
 \begin{figure*}[bthp]      
 \epsscale{0.9}	
 \plotone{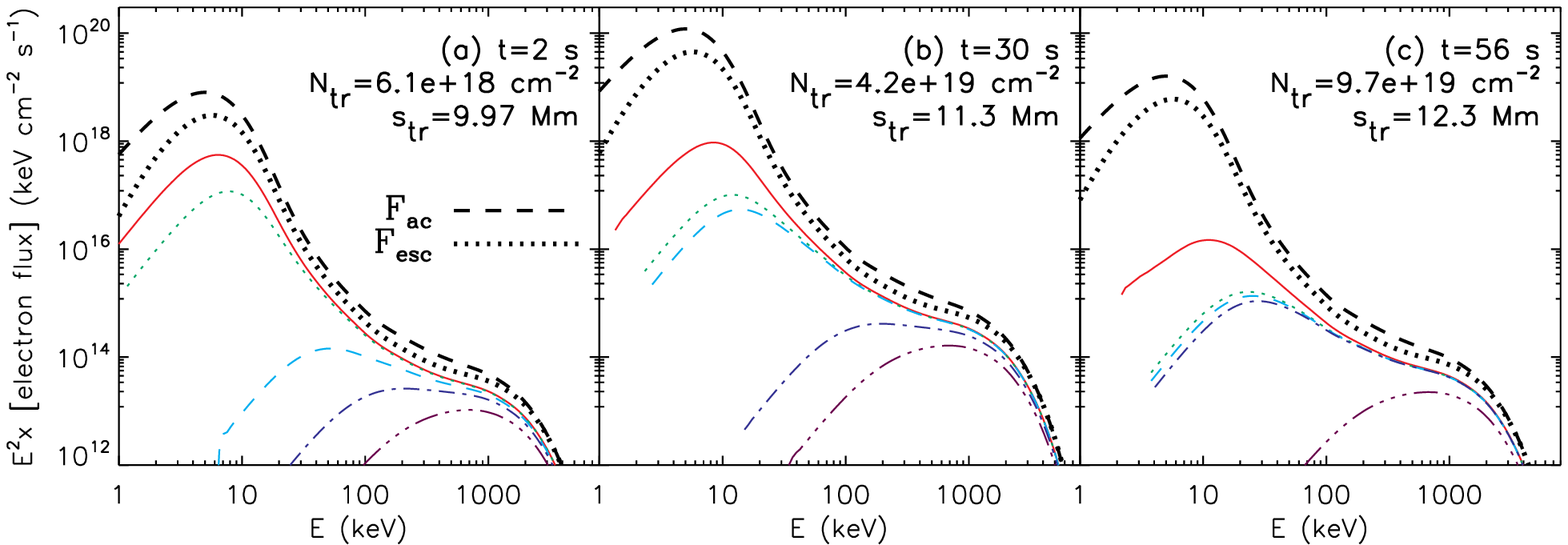} 
 \plotone{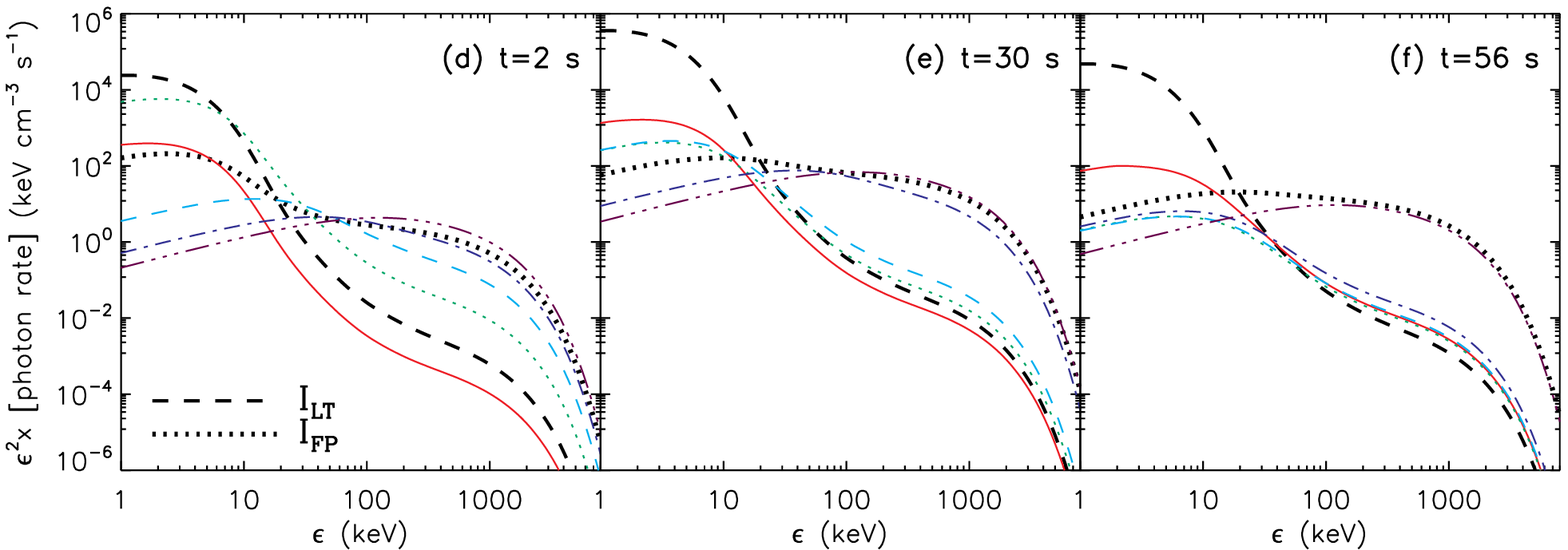} 
 \caption[]
 {Spectra of angle-integrated electron flux ({\it top}) and corresponding photon emission rate ({\it bottom}) 
 multiplied by energy squared at three selected times for Run~N. 
 The thick dashed line represents the LT (acceleration region) spectrum ($F_{\rm ac}$ or $I_{\rm LT}$), 
 while the thin colored lines ({\it solid}, {\it dotted}, {\it dashed}, {\it dot-dashed}, 
 and {\it triple-dot--dashed}) are the spectra at distances $s=$ 4, 10, 11, 12, and 13~Mm 
 from the injection site. The thick dotted line indicates the escaping electron flux ($F_{\rm esc}$) in the top panels
 and the equivalent FP photon emission rate ($I_{\rm FP}$) in the bottom panels. 
 The legends show the current values of the position ($s_{\rm tr}$) and column depth ($N_{\rm tr}$) of the TR.
 } \label{elphspec.eps}
 \end{figure*}
%
Early in the flare (Figs.~\ref{cmpr_Tv.eps}{\it a} and \ref{cmpr_Tv.eps}{\it c}),	
Run~H has higher upflow velocities and downflow temperatures than Run~N,
owing to its stronger net direct heating 	
with a deeper extent in the chromosphere (Fig.~\ref{cmpr_heating.eps}{\it c}). 
Run~N has an additional hot downflow component at $\gtrsim$2~MK,		
due to the expansion of the heated upper corona mentioned earlier. 
This hot component also dominates the VDEM at $v_{\rm LOS}<0$ (Fig.~\ref{cmpr_Tv.eps}{\it c})
because of the sharp rise of $G(T)$ up to $T=29 \MK$.
Later at $t=28 \s$ (Fig.~\ref{cmpr_Tv.eps}{\it b}),		
Run~N overtakes H in upflow velocity and temperate, while its hot coronal downflow has disappeared.
Run~N has a bimodal VDEM (Fig.~\ref{cmpr_Tv.eps}{\it d}) with a strong stationary, hot
component located near the loop apex (also see Fig.~\ref{HDmosaic_R.eps}), 
while Run~H exhibits a more gradual progression toward high velocities. 
This distinction is similar to what \citet[][see their Fig.~2]{Newton.VDEM.1995ApJ...447..915N} 
found for models with different initial coronal densities.

%

\section{Effects of Hydrodynamics on Particle Transport and X-ray Emission}		
\label{sect_el-ph}

We now turn our attention to the effects of fluid dynamics on particle characteristics, namely,
electron transport and nonthermal radiation. Here we present the
result of Run~N as a typical example. We will examine first the
energy distributions of electrons and bremsstrahlung photons, and then their spatial distributions.

\subsection{Electron and Photon Energy Spectra}
\label{subsect_elph-spec}

Figure~\ref{elphspec.eps} ({\it top panels}) shows the evolution of the angle-integrated 
electron flux spectrum $E^2 F(E, s)$ at different locations in the loop. 
In general, at a given time and a moderately large distance or column depth,
there is a deficit in low-energy electrons due to collisional energy losses and scatterings
on the paths of the electrons from the injection site. This appears as a turnover 	
in the spectrum and slight spectral hardening just above the turnover energy. 
As distance increases, progressively more low-energy electrons are lost, 
and thus the overall flux decreases while the spectral turnover shifts to higher energies 
\citep{LeachJ1981ApJ...251..781L}.
At $t=2 \s$ (Fig.~\ref{elphspec.eps}{\it a}) the TR is located
at distance $s_{\rm tr}=9.97 \Mm$ and column depth $N_{\rm tr}=6.1 \E{18} \pcms$.
The two fluxes at $s=4$ and 10~Mm ({\it thin solid}, {\it dotted}) are located in the
low-density corona or TR at small column depths from the injection site, and thus
appear similar in shape to the escaping flux $F_{\rm esc}$ ({\it thick dotted}). Other fluxes 
at $s=11$, 12, and 13~Mm are located in the chromosphere at large column depths, and thus
exhibit substantial reduction of low-energy electrons.

 \begin{figure*}[tbhp]      
 \epsscale{0.9}	
 \plotone{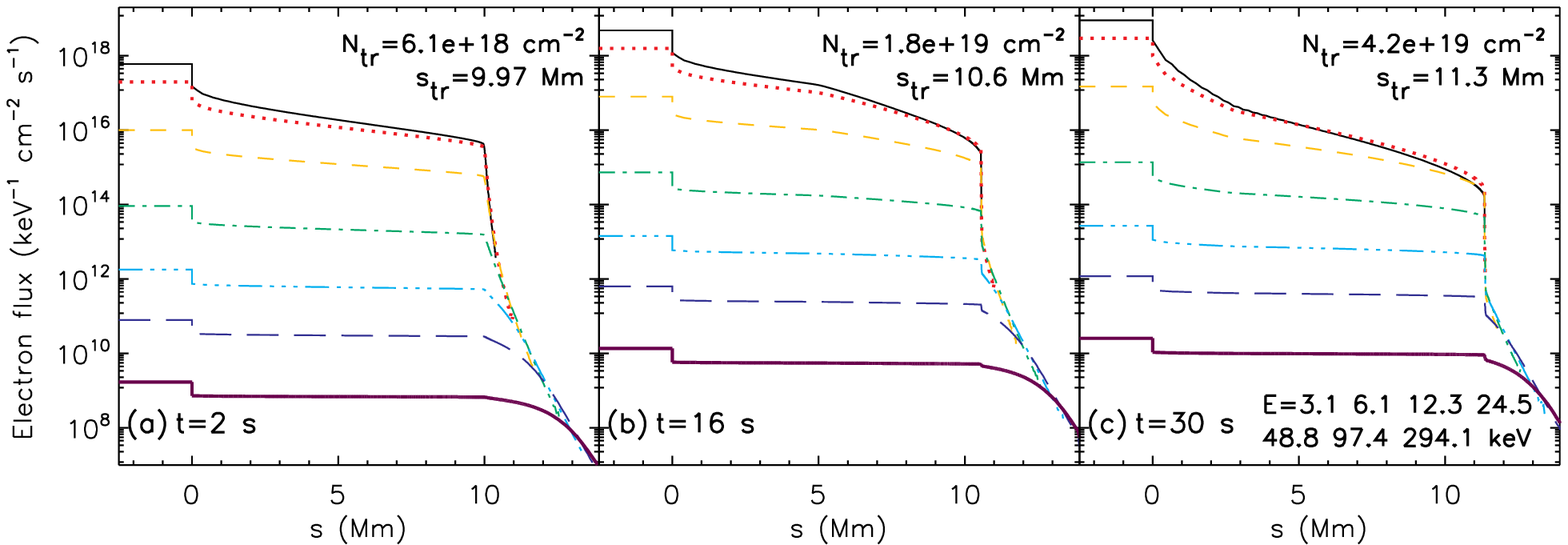} 
 \plotone{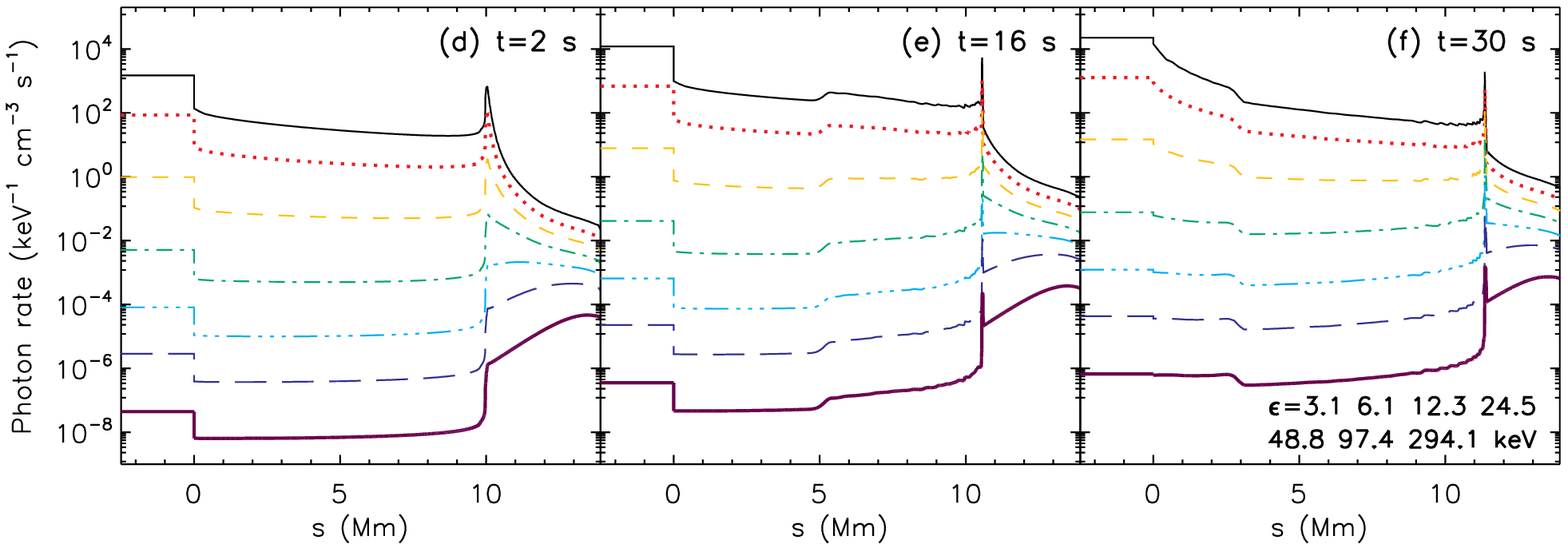} 
 \caption[]
 {Spatial distributions of angle-integrated electron flux ({\it top}) and photon emission rate  ({\it bottom})
 at three selected times for Run~N. From top to bottom, different lines in each panel
 represent electron or photon energies of 3.1, 6.1, 12.3, 24.5, 48.8, 97.4, and 294.1~keV.
 The step in the $s<0$  region corresponds to one half length of the acceleration region at
 the top of the loop (see Fig.~\ref{model-geometry.eps}). The step at the TR is
 due to the jump in the ambient density.
 } \label{elphdepth.eps}
 \end{figure*}
In this study, the flux ($F_{\rm esc}$) injected into the transport region 
does not change with time in spectral shape but only varies in normalization.
So does the electron flux at a given column depth. 
However, as chromospheric evaporation develops, the TR retreats to 
lower altitudes (Fig.~\ref{HDmosaic_R.eps}), while the coronal density increases. 
This causes the change of the column depth and thus the electron spectrum at 
a fixed location, which is what we notice here.
At $t=30 \s$ (Fig.~\ref{elphspec.eps}{\it b}) when the TR shifts down to $s_{\rm tr}=11.3 \Mm$ at 	
$N_{\rm tr}=4.2 \E{19} \pcms$, the spectrum at $s=11$~Mm		
looks similar to the other coronal spectra (at $s=4$ and 10~Mm) but different from the
chromospheric spectra.		
Meanwhile, the relative difference between the first coronal spectrum ($s=4$~Mm)
and the injected spectrum $F_{\rm esc}$ becomes larger, because of the increasing coronal
density and column depth between them. This trend continues through the end of the simulation.

The bottom panels of Figure~\ref{elphspec.eps} show the corresponding bremsstrahlung photon spectra 
$\epsilon^2 I(\epsilon, s)$ defined in equation~(\ref{nonthBremEq}).
Like its electron counterpart, the LT photon spectrum $\epsilon^2 I_{\rm LT}$ ({\it thick dashed})
shows a thermal-like component at low energies and a nonthermal tail 	
at intermediate energies with a cutoff at high energies.		
The equivalent FP spectrum $\epsilon^2 I_{\rm FP}$ below $s_{\rm tr}$ ({\it thick	dotted}, defined in \S\ref{subsect_brem})	
is harder than the LT spectrum. Their shapes in the 10s--100s keV range are
commonly observed for LT and FP sources. 
At larger distances, the spectra ({\it thin lines}) become progressively
harder because the corresponding electron spectra have the same variation \citep{LeachJ1983ApJ...269..715L}
as we have noticed above, and high-energy ($>$10s~keV) emission increases because of higher ambient densities.
As the coronal density increases with time, for the same reason mentioned above	
photon spectra at positions originally located in the corona
become progressively harder. At the same time, as the TR treats,
spectra at positions originally in the chromosphere but now in the corona become softer.
In other words, the difference in spectral shape between X-ray emissions at different
positions diminishes with time (Figs.~\ref{elphspec.eps}{\it e} and \ref{elphspec.eps}{\it f}).

\subsection{Electron and Photon Spatial Distributions}
\label{subsect_elph-depth}

Figure~\ref{elphdepth.eps} ({\it top panels}) shows the evolution of the angle-integrated
electron flux $F(E, s)$ as a function 
of distance at different energies from 3 to $300 \keV$. The step at $s=0$ is owing to
the difference between the LT (acceleration region) flux $F_{\rm ac}$ and the escaping flux $F_{\rm esc}$
mentioned before (see Fig.~\ref{init_elspec.eps}).
In general, the electron flux decreases
with distance or column depth from the injection site. The slope of the curve
$-dF(E, s)/ds$ is steeper at lower energies because lower energy electrons 
lose energy faster and are more sensitive to pitch-angle scattering. At a given energy, 
the slope depends on the ambient density, because 
$dF(E, s)/ds = n_e [dF(E, N)/dN]$,		
where $F(E, N)$ is generally a smooth function of $N$ \citep{McTiernanJ1990ApJ...359..524M}.
Therefore, if there is a rapid increase in density with distance, the slope increases
quickly; the opposite would be true if the density were to decrease sharply. 
A sharp break is obvious at the TR, and a milder break occurs at the evaporation front
(e.g., at $s\simeq 5 \Mm$ in Fig.~\ref{elphdepth.eps}{\it b}). 	
An upward break or flattening is visible at $s\simeq 3 \Mm$ where
the density jump is reflected back from the loop apex at $t=30 \s$ (Fig.~\ref{elphdepth.eps}{\it c}). 
It is also visible just below the TR (see Figs.~\ref{elphdepth.eps}{\it b} and \ref{elphdepth.eps}{\it c}),
where a sharp density decrease from the narrow density spike occurs (Fig.~\ref{HDmosaic_R.eps}).
As density increases 	
due to chromospheric evaporation, the spatial distribution in the corona becomes steeper.
The slope variation with distance (i.e., breaks) and time is more pronounced
at lower energies for the same reason noted above.

Figure~\ref{elphdepth.eps} ({\it bottom panels}) shows the evolution of the 
corresponding spatial distribution of angle-integrated bremsstrahlung emission $I(\epsilon, s)$.	
In the early stage, low-energy emission comes primarily from the LT, 
while high-energy emission is concentrated below the TR.
Because nonthermal bremsstrahlung radiation is proportional to both the electron flux 
and local target density, photon emission reflects details of the density distribution 
in a more pronounced way than the electron flux profile.	
As is evident, the emission profile clearly indicates	
density features, including the evaporation front and the density spike just below the TR.
The local emission enhancement at the evaporation front moves upward with time,
which may be responsible for the observed X-ray sources moving along
the loop \citep{LiuW2006ApJ...649.1124L, SuiL2006ApJ...645L.157S} mentioned in \S~\ref{subsect_intro1}.
As time proceeds, relatively more emission comes from the coronal portion of the loop
because of the increased density. Specifically, at low energies, the emission intensity
decreases with distance in the corona more sharply than before. 
At intermediate energies, we find a temporal transition from FP-dominated emission to
LT-dominated emission, which occurs at progressively higher energies,
reminiscent of the phenomenon observed by \citet{LiuW2006ApJ...649.1124L}. 
At high energies, such a change is not visible	
because the high density corona still looks transparent to high-energy electrons,
but the retreat of the TR is obvious. 

\section{Summary and Discussion}
\label{sect_conclude}

We have performed simulations of solar flares that self-consistently combine
acceleration, transport, and radiation of energetic electrons 
(using the Stanford unified code) with fluid dynamics of the atmospheric response 
(using the NRL flux tube code).	
As the first successful one of its kind, this model improves on previous HD
simulations in two major aspects. First, it includes more accurate evaluation of electron heating 
from full Fokker-Planck calculation of particle transport. Second, it uses 
a more realistic electron spectrum from the stochastic acceleration model (PL04) as 
the injection to the transport region.	
We compare this more advanced treatment with  	
models in which an {\it ad hoc} electron distribution
of a power-law with a low-energy cutoff is injected into the loop and/or transport
is dealt with approximately. Our conclusions are:

 1.~For the specific injection of beamed, power-law electrons,	
the old analytical model of MEL89 (Run~O) provides an acceptable approximation.
Its result differs by $\sim$10\% from that of the reference hybrid model (Run~H) obtained by
the more accurate Fokker-Planck calculation		
(see Table~\ref{table_cases} and Figs.~\ref{cmpr_history.eps} and \ref{HDcmpr_multi.eps}).



 2.~In the new model (Run~N), where the injected electron spectrum is based on stochastic
acceleration, we find higher coronal temperatures and densities,
larger upflow velocities, and faster increases of these quantities than the
hybrid model (Run~H,	
Fig.~\ref{cmpr_history.eps}). This is mainly because
the new injected electron spectrum smoothly spans from a quasi-thermal
component to a nonthermal tail (Fig.~\ref{init_elspec.eps}). The low-energy electrons in the quasi-thermal regime,
which contain the bulk of the total energy budget, deposit their energy
primarily in the corona. This results in significant coronal heating and thus a large
downward heat conduction flux that helps drive ``evaporation" of plasma at the TR. 
In contrast, the electron spectrum in the hybrid model with a low-energy cutoff leads
to more energy directly deposited in the chromosphere, which is radiated away more quickly,
leaving less energy to produce actual heating (Figs.~\ref{HDcmpr_multi.eps} and \ref{cmpr_heating.eps}).
This is qualitatively consistent with the conclusion of MEL89 where an electron spectrum with a smaller
low-energy cutoff or steeper slope resulted in a stronger chromospheric evaporation.

 3.~The energy and spatial distributions of energetic electrons and bremsstrahlung photons
bear the fingerprint of the changing density distribution caused by chromospheric
evaporation. In general, as time proceeds, the electron and photon spectra at
positions remaining in the corona become progressively harder because of the increasing coronal density,
while those at positions previously in the chromosphere and now in the corona
(due to the retreat of the TR) become softer (Fig.~\ref{elphspec.eps}). Any density
jump in space results in a sudden change in the spatial distributions of
energetic electrons and X-ray photons (Fig.~\ref{elphdepth.eps}). In particular, the evaporation front appears
as a local emission enhancement, which, in principle, can be imaged by X-ray telescopes.


\subsection{Comparison with Observations}	
\label{subsect_future-obs}

Over several decades \citep[see review by][]{AntonucciE1999ManyFacesOfTheSun}, Doppler observations have indicated	
hot, fast ($\gtrsim$$100 \km \ps$) upflows \citep{DoschekG1980ApJ...239..725D}		
and cool, slow ($\gtrsim$$10 \km \ps$) downflows \citep{WuelserJ1994ApJ...424..459W, BrosiusJ2004ApJ...613..580B}
during flares. This is consistent with chromospheric evaporation and momentum recoil 
shown in our and earlier simulations \citep[e.g.,][]{MariskaJ1982ApJ...255..783M, FisherG1985ApJ...289..414F}. 
In addition, \citet{MilliganDennis.2009ApJ.EIS-v-T}	
reported plasma velocities at multiple temperatures obtained from {\it Hinode} EUV Imaging Spectrometer (EIS) 
observations, which show excellent agreement with the Run~N curve 
in Figure~\ref{cmpr_Tv.eps}{\it b} throughout the $10^4$--$10^7$~K range.
Such an agreement with HD simulations has not been seen before.

Heat conduction, compared with more popular direct electron heating, 
plays an important role in driving chromospheric evaporation in our new model.
Observational support of this 	
was first reported by \citet{ZarroLemen.ConductionDriven.1988ApJ...329..456Z}, 
who found an upflow velocity $\lesssim 50 \km \ps$ at $T\simeq 6 \MK$ during the cooling phase of a flare.
\cite{MilliganR.hotMicro.2008ApJ...680L.157M} recently reported an unusually high temperature (2~MK)	
downflow at $\sim$$14 \km \ps$ in a microflare with no detection of HXRs (implying a low flux of nonthermal electrons),
which possibly results from conductive heating.  This downflow could be due to the thermal expansion 
early in the corona (Fig.~\ref{cmpr_Tv.eps}{\it a}) or 
the momentum recoil later in the chromosphere 		
(Fig.~\ref{cmpr_Tv.eps}{\it b}). More recently, \citet{Battaglia.conduct-evapor_2009A&A} 
interpreted the growths of the SXR emission measure observed early during flares (before HXRs being observed) 
as new evidence of conduction-driven chromospheric evaporation.

%
Our simulations predict that evaporation upflows tend to have higher
temperatures when conductive heating dominates over direct electron heating, 
while the opposite is true for recoil downflows (Fig.~\ref{cmpr_Tv.eps}).
This can be checked against observed distributions of the plasma bulk velocity vs.~temperature
\citep[e.g.,][]{MilliganDennis.2009ApJ.EIS-v-T}.
Our simulated VDEM can also be readily used to synthesize emission lines
\citep{Newton.VDEM.1995ApJ...447..915N} to be compared with observations
and help differentiate theoretical models, because of the sensitive dependence of VDEM on heating mechanisms.




\subsection{Future Work}	
\label{subsect_future-num}

This paper is the first in a series and we have established the numerical model and presented
initial results. In followup papers, we will explore the parameter space and use this model to 
investigate the Neupert effect and the observed moving X-ray sources 
\citep{LiuW2006ApJ...649.1124L, SuiL2006ApJ...645L.157S}. 
More importantly, we will incorporate the atmospheric feedback on the acceleration process.
This is because chromospheric evaporation may change the physical condition
(e.g., plasma density and temperature) in the LT acceleration region.
The density enhancement, for example, causes the ratio of electron plasma frequency to
gyro-frequency $\alpha= \omega_{\rm pe}/\Omega_e \propto n_e^{1/2}/B_0$ to increase.
This can lead to the reduction of the efficiency of electron acceleration (PL04)
and thus the quenching or spectral softening \citep[e.g.,][]{ParksWinckler.SHS.1969ApJ...155L.117P}
of nonthermal HXR tails observed during the late stages of flares.


Some technical aspects of this model can be improved in the future:
 (1) The fully ionized hydrogen plasma assumed here can be replaced by a plasma of a 
more realistic solar abundance, with the inclusion of neutrals and ionization equilibrium.
 (2) The ``cold" target assumption in the transport code can be abandoned and replaced
with Coulomb diffusion in energy space \citep{SpitzerL1962pfig.book.....S, MillerJ1996ApJ...461..445M}
for a general ``warm" target plasma \citep{Emslie2003ApJ...595L.119E}.
 (3) In particle transport
 \footnote{Stochastic acceleration of ions has been modeled 
 \citep{Miller.proton.1995ApJ...452..912M, PetrosianV2004ApJ...610..550P, 
 LiuS2004ApJ...613L..81L, LiuS2006ApJ...636..462L} and implemented in the Stanford unified code.
 } 
and HD calculations, one can include energetic protons \citep[c.f.,][]{Emslie.proton.1998ApJ...498..441E}
and heavier ions, whose momentum loss to the background plasma, 
in addition to the overpressure \citep{KosovichevA2006SoPh} produced by electron heating, could
contribute to generating seismic waves observed in some flares
\citep{KosovichevZharkova1998Natur, Donea.Lindsey.2005ApJ...630.1168D}.
(4) In the long run, we intend to implement time-dependent particle transport
calculation, full loop simulation in an asymmetric geometry, return currents, and radiative transfer.


The combined treatment of the particle and fluid descriptions of plasma presented here
opens a door to a broad range of applications.	
This model, with proper modifications, can be applied to environments where interrelated
particle acceleration and transport and plasma flows are present,	
such as (exo)planetary auroras \citep[e.g.,][]{LiuAirapetian2008AAS...21115901L} and
flares on other stars and in accretion disks near black holes.


\acknowledgements
{This work was primarily supported by NASA grants NAG5-12111, NAG5 11918-1, and NSF grant ATM-0312344
at Stanford University. Writing of this paper was partially conducted during W.~Liu's appointment
to the NASA Postdoctoral Program at the Goddard Space Flight Center,
administered by Oak Ridge Associated Universities.	
He is grateful to his postdoctoral advisers, Brian Dennis and Gordon Holman, for fruitful discussions.
Work performed by J.~Mariska was supported by NRL basic research funds.
The authors thank the referee for constructive comments and
many individuals, including S.~Liu, W.~East, T.~Donaghy, J.~Pryadko,
B.~Park, R.~Hamilton, J.~McTiernan, and J.~Leach, who contributed 
to the Stanford unified Fokker-Planck code over three decades.
} 





{\scriptsize
\bibliography{bib/ads_all_edit,bib/LiuW-group,bib/Liu-Wei}
}

\end{document}